\documentclass[12pt]{article}
\usepackage{amsmath}
\usepackage{amssymb}
\usepackage{graphicx,psfrag,epsf}
\usepackage{enumerate}
\usepackage{natbib}
\usepackage{bm}
\usepackage{subfig}
\usepackage{bbm}
\newcommand{\blind}{0}
\usepackage{multirow}
\usepackage{float}
\usepackage[margin=1in]{geometry}
\usepackage{caption}
\usepackage{color}
\usepackage[dvipsnames]{xcolor}
\usepackage{setspace}

\allowdisplaybreaks

\begin{document}

\def\spacingset#1{\renewcommand{\baselinestretch}%
{#1}\small\normalsize} \spacingset{1}
\newcommand{\bb}{{\bf b}}
\newcommand{\bd}{{\bf d}}
\newcommand{\bv}{{\bf v}}
\newcommand{\bx}{{\bf x}}
\newcommand{\by}{{\bf y}}
\newcommand{\bw}{{\bf w}}
\newcommand{\bz}{{\bf z}}
\newcommand{\bh}{{\bf h}}
\newcommand{\bc}{{\bf c}}
\newcommand{\bff}{{\bf f}}
\newcommand{\bu}{{\bf u}}
\newcommand{\be}{{\bf e}}
\newcommand{\ba}{{\bf a}}
\newcommand{\bi}{{\bf i}}
\newcommand{\bt}{{\bf t}}
\newcommand{\bs}{{\bf s}}

\newcommand{\bca}{{\bf A}}
\newcommand{\bce}{{\bf E}}
\newcommand{\bcb}{{\bf B}}
\newcommand{\bcd}{{\bf D}}
\newcommand{\bcg}{{\bf G}}
\newcommand{\bch}{{\bf H}}
\newcommand{\bci}{{\bf I}}
\newcommand{\bcm}{{\bf M}}
\newcommand{\bcr}{{\bf R}}
\newcommand{\bct}{{\bf T}}
\newcommand{\bcv}{{\bf V}}
\newcommand{\bcy}{{\bf Y}}
\newcommand{\bcz}{{\bf Z}}
\newcommand{\bcx}{{\bf X}}
\newcommand{\bcw}{{\bf W}}
\newcommand{\bcf}{{\bf F}}
\newcommand{\bck}{{\bf K}}
\newcommand{\bcl}{{\bf L}}
\newcommand{\bcu}{{\bf U}}
\newcommand{\bcs}{{\bf S}}

\def\tilde{\widetilde}

\newcommand{\ita}{\eta}
\newcommand{\bita}{\bm{\eta}}
\newcommand{\bzeta}{\bm{\zeta}}
\newcommand{\balpha}{\bm{\alpha}}
\newcommand{\bbeta}{\bm{\beta}}
\newcommand{\bfdelta}{\bm{\delta}}
\newcommand{\bfDelta}{\bm{\Delta}}
\newcommand{\bfeta}{\bm{\eta}}
\newcommand{\bGamma}{\bm{\Gamma}}
\newcommand{\bgamma}{\bm{\gamma}}
\newcommand{\blambda}{\bm{\lambda}}
\newcommand{\bLambda}{\bm{\Lambda}}
\newcommand{\bmu}{\bm{\mu}}
\newcommand{\bfomega}{\bm{\omega}}
\newcommand{\bomega}{\bm{\Omega}}
\newcommand{\bphi}{\bm{\Phi}}
\newcommand{\bPsi}{\bm{\Psi}}
\newcommand{\bpsi}{\bm{\psi}}
\newcommand{\bspsi}{\bm{\psi}}
\newcommand{\bsigma}{\bm{\Sigma}}
\newcommand{\btau}{\bm{\tau}}
\newcommand{\btheta}{\bm{\theta}}
\newcommand{\beps}{\bm{\varepsilon}}
\newcommand{\bfphi}{\bm{\varphi}}
\newcommand{\bxi}{\bm{\xi}}
\newcommand{\bpi}{\bm{\pi}}
\newcommand{\bepsilon}{\bm{\epsilon}}

\newcommand{\bzero}{{\bf 0}}
\newcommand{\bone}{{\bf 1}}
\newcommand{\Ind}{\text{I}}
\newcommand{\logit}{\text{logit}}
\newcommand{\rank}{\text{rank}}
\newcommand{\diag}{\text{diag}}
\newcommand{\for}{\text{for}}
\newcommand{\de}{\stackrel{D}{=}}
\newcommand{\te}{\stackrel{\Delta}{=}}
\newcommand{\ind}{\stackrel{ind}{\sim}}
\newcommand{\vs}{\vspace{0.1in}}
\newcommand{\V}{\mathbf V}
\newcommand{\I}{\mathbf I}
\newcommand{\bo}{\boldsymbol }

\newtheorem{theorem}{Theorem}
\newtheorem{proposition}{Proposition}
\newtheorem{corollary}{Corollary}
\newtheorem{assumption}{Assumption}
\newtheorem{definition}{Definition}


\if0\blind
{
  \title{
  A Blockwise Mixed Membership Model for Multivariate Longitudinal Data: Discovering Clinical Heterogeneity and Identifying Parkinson's Disease Subtypes
  }
  \author{Kai Kang\thanks{
   kangk5@mail.sysu.edu.cn}\hspace{.2cm}\\
   Department of Statistics, Sun Yat-sen University\\
   and \\
   Yuqi Gu\thanks{
   yuqi.gu@columbia.edu}\\
   Department of Statistics, Columbia University}
  \date{}
  \maketitle
} \fi

\if1\blind
{
  \bigskip
  \bigskip
  \bigskip
  \begin{center}
    {\LARGE\bf Title}
\end{center}
} \fi

\begin{abstract}
Current diagnosis and prognosis for Parkinson's disease (PD) face formidable challenges due to the heterogeneous nature of the disease course, including that (i) the impairment severity varies hugely between patients, (ii) whether a symptom occur independently or co-occurs with related symptoms differs significantly, and (iii) repeated symptom measurements exhibit substantial temporal dependence. To tackle these challenges, we propose a novel blockwise mixed membership model (BM$^3$) to systematically unveil between-patient, between-symptom, and between-time clinical heterogeneity within PD. The key idea behind BM$^3$ is to partition multivariate longitudinal measurements into distinct blocks, enabling measurements within each block to share a common latent membership while allowing latent memberships to vary across blocks. Consequently, the heterogeneous PD-related measurements across time are divided into clinically homogeneous blocks consisting of correlated symptoms and consecutive time. From the analysis of Parkinson’s Progression Markers Initiative data ($n=1,531$), we discover three typical disease profiles (stages), four symptom groups (i.e., autonomic function, tremor, left-side and right-side motor function), and two periods, advancing the comprehension of PD heterogeneity. Moreover, we identify several clinically meaningful PD subtypes by summarizing the blockwise latent memberships, paving the way for developing more precise and targeted therapies to benefit patients. Our findings are validated using external variables, successfully reproduced in validation datasets, and compared with existing methods. Theoretical results of model identifiability further ensures the reliability and reproducibility of latent structure discovery in PD.

\end{abstract}

\noindent%

\spacingset{1.7}

\section{Introduction}
\subsection{Background}
Parkinson's disease (PD) is one of the most common neurodegenerative disorders and characterized by a series of clinical symptoms, including but not limited to bradykinesia, rigidity, sleep disorders, cognitive impairment,
depression, and dementia. However, there is substantial variability in PD clinical presentation. For example, some individuals demonstrate tremor as the dominant and persistent motor feature from the onset of their disease, whereas others initially show cognitive impairment, neuropsychiatric symptoms and autonomic dysfunction \citep{tolosa2009diagnosis}. In addition, although most patients with PD eventually develop movement disorder, a minority do not and even among those PD patients who develop motor symptoms, the pace varies widely \citep{pigott2015longitudinal}. Indeed, the timescale over which patients may develop movement disorder ranges from months to years to decades. 

Investigating the aforementioned complex clinical heterogeneity of PD is critical for understanding the underlying disease process, developing better therapies and disease management. 
By dividing patients into small subtypes with common features, neurologists can investigate key differences in the underlying pathological processes. Once we have a better understanding of the biological basis of these subtypes, we can begin to develop targeted treatments \citep{greenland2019clinical}. In addition, we can use our knowledge regarding clinical heterogeneity to improve the disease management. For example, an individual presenting with isolated motor symptoms and newly diagnosed with PD will have different needs
from a PD patient with cognitive dysfunction, who may require social or family support to perform activities of daily living. Furthermore, exploration into the heterogeneity of PD facilitates clinicians to select the most appropriate patients for inclusion in clinical trials.

This study is motivated from Parkinson's Progression Markers Initiative (PPMI), a large international study of PD that aims to identify biological markers of Parkinson’s risk, onset and progression. With the goal of capturing the full clinical course of the disease, PPMI collects multimodality data from longitudinal follow-up of various cohorts.  However, the complex multivariate longitudinal data makes the investigation of clinical heterogeneity even more challenging. We will thoroughly discuss the three main challenges in discovering the clinical heterogeneity in PPMI study in Section \ref{sec:1.2}. 

\subsection{Challenges in the Motivating PPMI Dataset}
\label{sec:1.2}
There are at least three challenges in discovering the clinical heterogeneity of PD in PPMI study. 
First, during the disease course, subjects exhibit substantial between-individual heterogeneity measured by multiple clinical variables. For instance, some PD patients suffer from severe sleeping disturbance while others only have mild symptoms. Based on the functional deficits (disability) and objective signs (impairment), \cite{hoehn1967parkinsonism} first designed the five-point scale, Hoehn and Yahr scale, to partition patients into subtypes and describe how motor symptoms progress in PD. However, such a scale has only five options and therefore a large variety of impairment severities is collapsed together. Over the last decades, data-driven clustering techniques, including but not limited to K-means \citep{macqueen1967some}, latent class model \citep{goodman1974}, and finite mixture model \citep{peel2000finite} have been used to address the between-individual heterogeneity in PD. These methods assume the existence of a single true clustering in a dataset and assign each subject a single subtype that is defined by all PD-related symptoms. Nevertheless, given that PD is usually multifaceted and can be meaningfully partitioned in multiple ways \citep{thenganatt2014parkinson,marras2015subtypes}, a single generic subtype does not hold for PD. Thus, it is desirable to assign multiple clusters for a subject, where each cluster reveals the subtype of the specific aspect of PD. 

Second, in addition to individual heterogeneity, significant between-symptom heterogeneity also exists in PD and brings us a second challenge in the analysis of PD. For example, Figure \ref{fig:Cramer's V}D showcases the sample correlation of 29 common motor and nonmotor clinical symptoms in the PPMI study. It is apparent that the dependence among NP3NIGLU (Rigidity-left upper extremity), NP3NIGLL (Rigidity-left lower extremity) and NP3PRSPL (Pronation-supination-left hand) are relatively strong whereas that between NP3NIGLU and NP2TRMR (tremor) is weak. Such a dependence structure illustrates that subjects' clinical symptoms are not homogeneous across different domains, for example some patients may present tremor but have not experienced postural instability. Previous studies have also repeatedly found distinct PD subgtypes, such as tremor dominant, nontremor dominant or postural instability gait disorder phenotype, further confirming the existence of between-symptom heterogeneity. As a result, naively assuming symptoms are all correlated (see Figure \ref{fig:Cramer's V}A) or all independent (see Figure \ref{fig:Cramer's V}B) cannot capture the subtle between-symptom heterogeneity and fails to recover the dependence structure among PD-related biomarkers. 

Third, to understand the progression of PD, subjects enrolled in PPMI study will undergo a longitudinal schedule of clinical assessment at screening/baseline and at 3 month intervals during the first year of participation and then every 6 months thereafter. Given that subjects' clinical symptoms measured in adjacent time points tend to share certain similarities (e.g., frailty, disease status), time dependence is the third challenge and should be addressed when modeling disease progression through repeated measurements of clinical symptoms. \cite{severson2020personalized} developed a hidden Markov model for learning the serial dependence of motor function and disease progression of PD. \cite{wang2024multilayer} combined the linear trend and the sigmoidal trajectory to learn the change-point where a subject's disease status transiting from normal to severe. However, these methods either focus on Gaussian data, ignore the correlation between symptoms, or assume subject-wise homogeneity. Simultaneously accounting for between-individual, between-symptom and between-time heterogeneity in multivariate longitudinal PD symptoms remains a challenge.

\subsection{Our Contributions}
To the best of our knowledge, our proposed approach is the first to systematically discover the between-patient, between-symptom, and between-time clinical heterogeneity of PD. Specifically, we propose a {\it blockwise mixed membership model} (BM$^3$) for multivariate longitudinal biomarkers measured in the PPMI study. Through the subject-specific mixed membership score, BM$^3$ permits each subject to belong to multiple clusters, thus facilitating the characterization of the multifaceted disease in diverse ways and addressing the individual heterogeneity. 
Grade of Membership models (GoMs) \citep{woodbury1978gom,erosheva2004mixed, erosheva2007describing,manrique2014longitudinal}  are mixed membership models \citep{airoldi2015handbook} that allow each subject to be partial memberships of multiple different latent extreme profiles and largely extend the flexibility of simple mixture models or latent class models \citep{lazarsfeld1968latent,hagenaars2002applied}.
Different from these conventional GoMs,
the proposed BM$^3$ partitions symptoms and time points into several blocks, such that the observations within a block share a same latent membership while those across blocks can have different memberships. By doing so, the between-symptom and between-time correlation among measurements of relevant clinical symptoms in adjacent visits are well captured via the assigned same latent membership. Importantly, the BM$^3$ does not require the block structure (i.e., which observation belongs to which block) to be specified a priori, but fully determined by data. We propose a Bayesian method together with an efficient Markov Chain Monte Carlo (MCMC) algorithm to infer the unknown block structure, estimate parameters, and perform model selection. 

Based on the BM$^3$, we first aim to identify several typical disease profiles reflecting different levels of symptom severity. We expect to identify clinically meaningful symptom groups (also called syndromes) and time periods via the estimated block structure. Combining the blockwise latent memberships, we aim to partition patients into clinically meaningful and biologically homogeneous PD subtypes. Theoretically, we prove the BM$^3$ is strictly and generically identifiable under mild and easy-to-check conditions, ensuring that the model
parameters can be uniquely recovered from the observables. 
It is worth emphasizing that the block structure of how to group the symptoms and time points, is also identifiable. This provides the crucial theoretical guarantees for learning these meaningful structures from real data.
Practically, the proposed BM$^3$ provides a powerful method that can be easily modified and applied to other diseases (e.g., Alzheimer's disease) for investigating the complex heterogeneity in disease progression. 
The uncovered heterogeneity will provide an interpretable basis for downstream tasks such as designing personalized treatments or interventions.

The article is organized as follows. Section 2 introduces the essential block structure and the BM$^3$ for multivariate longitudinal categorical data. Section 3 proposes the identifiability conditions for BM$^3$. Section 4 specifies the prior distributions and develops the method for posterior inference. In Section 5, numerous simulation studies are conducted to evaluate the performance of the Bayesian methods in parameter estimation and model selection. Section 6 presents the analyses of the PPMI dataset and compare the result using BM$^3$ with those using conventional models. Section 7 concludes the paper.

\section{Blockwise Mixed Membership Model}
\label{sec:intro}
\subsection{Notation and Setup}
We begin by introducing the key notation. The multivariate longitudinal data were collected using the Movement Disorder Society-sponsored revision of the unified Parkinson’s disease rating scale \citep[MDS-UPDRS,][]{goetz2008movement}, a comprehensive assessment for monitoring the burden of PD in terms of motor and nonmotor experiences in daily life. For a positive integer $M$, denote $[M]=\{1,2,\ldots,M\}$.
Let $y_{i,j,t}$ denote the $i$'th subject's response to item $j\in[p]$ ($j$'th PD-related symptom) at time $t\in[T_i]$, where $p$ is the total number of MDS-UPDRS items, $n$ is the total number of subjects and $T_i$ is the number of visits for subject $i$. Throughout the paper, we will refer to PD-related item $j$ as ``symptom $j$" . Each observation $y_{i,j,t} \in \{1,\hdots,d_j\}$ takes one of $d_j$ unordered categories, reflecting the severity of symptom $j$. Let $K$ denote the number of extreme profiles. The $K$-dimensional probability simplex is defined as $\Delta^{K-1} = \{(\pi_1,\hdots,\pi_K):\pi_k \geq 0$ for all $k$, $\sum_{k=1}^{K} \pi_k =1 \}$. Each subject $i$ has an individual proportion vector $\bpi_i = (\pi_{i1},\hdots,\pi_{iK}) \in \Delta^{K-1}$, which indicates
the degrees to which subject $i$ partially belongs to each of the $K$ extreme latent profiles (i.e., latent classes).

\subsection{Key Block Structure}\label{sec:block}
\begin{figure}[h!]
\centering
            \includegraphics[height=0.4\textheight]{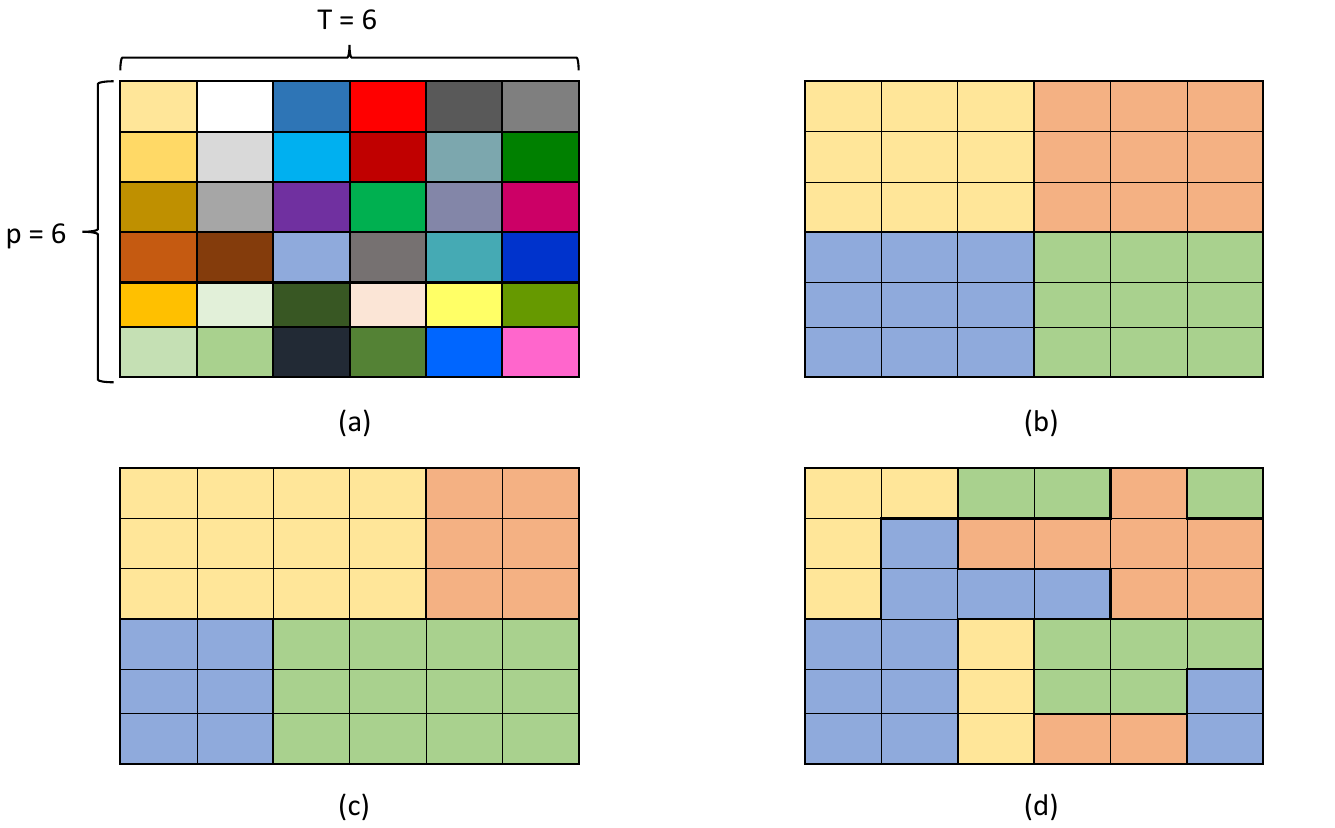}
            \caption{Block structure of multivariate longitudinal data with $p=6$ symptoms and $T=6$ time points. (a) Conventional GoM with $G=6$ and $R=6$. (b) BM$^3$ with homogeneous periods $\bb$, $G=2$ and $R=2$. (c) BM$^3$ with symptom group-specific periods $\bb_g$, $G=2$ and $R=2$. 
            (d) General BM$^3$ with $4$ overall blocks.} 
            \label{fig:block}
\end{figure}

To simplify the presentation, we initially assume that all  subjects share the same number of visits, $T_i=T$. For the $p\times T$ multivariate longitudinal observations $y_{i,j,t}$, where $j\in [p]$ and $ t\in[T]$, collected from subject $i$, conventional GoMs in \cite{erosheva2007describing} assign a latent membership $z_{i,j,t}$ to each observation (see Figure \ref{fig:block}a). Instead, the key idea of the proposed BM$^3$ is to further introduce a higher-level block structure, imposing useful constraints on the conventional GoMs.   Specifically, the $p$ symptoms and $T$ time points are partitioned into $G$ groups $(G\leq p)$ and $R$ periods $(R\leq T)$, respectively. Consequently, the total $p \times T$ multivariate longitudinal observations are divided into $G\times R$ {\it{blocks}}. Unlike GoMs, which assign each observation a latent membership, BM$^3$ assumes that the latent membership of observations are the same within a block but can differ across blocks. To formalize this block structure, we introduce notations of the group indicator $s_{j}$ and period indicator $b_{t}$. Let $\bs=(s_1,\hdots,s_p)$, where $s_j=g$ if and only if the $j$-th variable belongs to group $g$. Similarly, denote $\bb = (b_1,\hdots,b_T)$, where $b_{t}=r$ if and only if $t$-th time point belongs to the period $r$. Here we assume $b_{t}\leq b_{t+1}$ since only consecutive time points may belong to the same period in disease course. Figure \ref{fig:block}(b) illustrates an example of the block structure in BM$^3$ with $p = 6, T=6, G =2, R =2, \bs=(1,1,1,2,2,2)$ and $\bb=(1,1,1,2,2,2)$. 
Based on this block structure, the distribution of the observed $\{y_{i,1,1}\hdots,y_{i,p,T}\}$ in BM$^3$ can be expressed as
\begin{align}
p(y_{i,1,1},\hdots,y_{i,p,T}) = \int_{\Delta_{K-1}} \prod_{g=1}^G \prod_{r=1}^R \Big[\sum_{k=1}^K \pi_{i,k} \prod_{j:s_j=g} \prod_{t:b_t=r} 
f(y_{i,j,t} \mid \blambda_{j,k})
\Big] dD_{\balpha}(\bpi_i),
\end{align} 
where $\bpi_i$ follows some distribution $D_{\balpha}$ on the probability simplex, such as a  Dirichlet distribution, and $\balpha$ represents the population parameters for this distribution. 
Each symptom $j$ corresponds to a set of $K$ conditional distributions indexed by 
the parameter vector $\blambda_{j,k}$, which is symptom-and-extreme profile-specific. 

\begin{figure}[h!]
    \centering
    \subfloat[\centering Conventional GoM]
    {{\includegraphics[width=8cm]{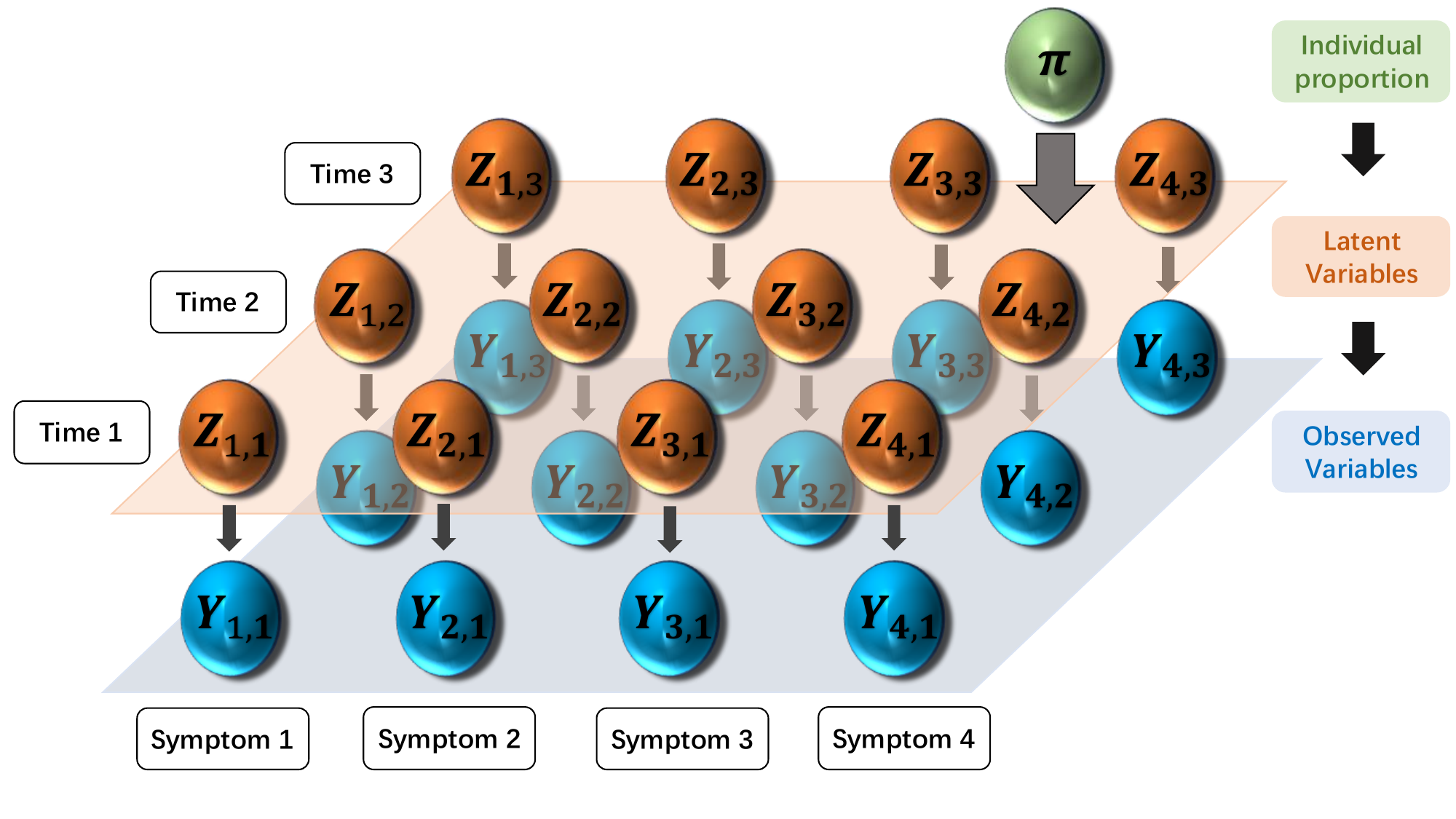} }}
    \subfloat[\centering BM$^3$]
    {{\includegraphics[width=8cm]{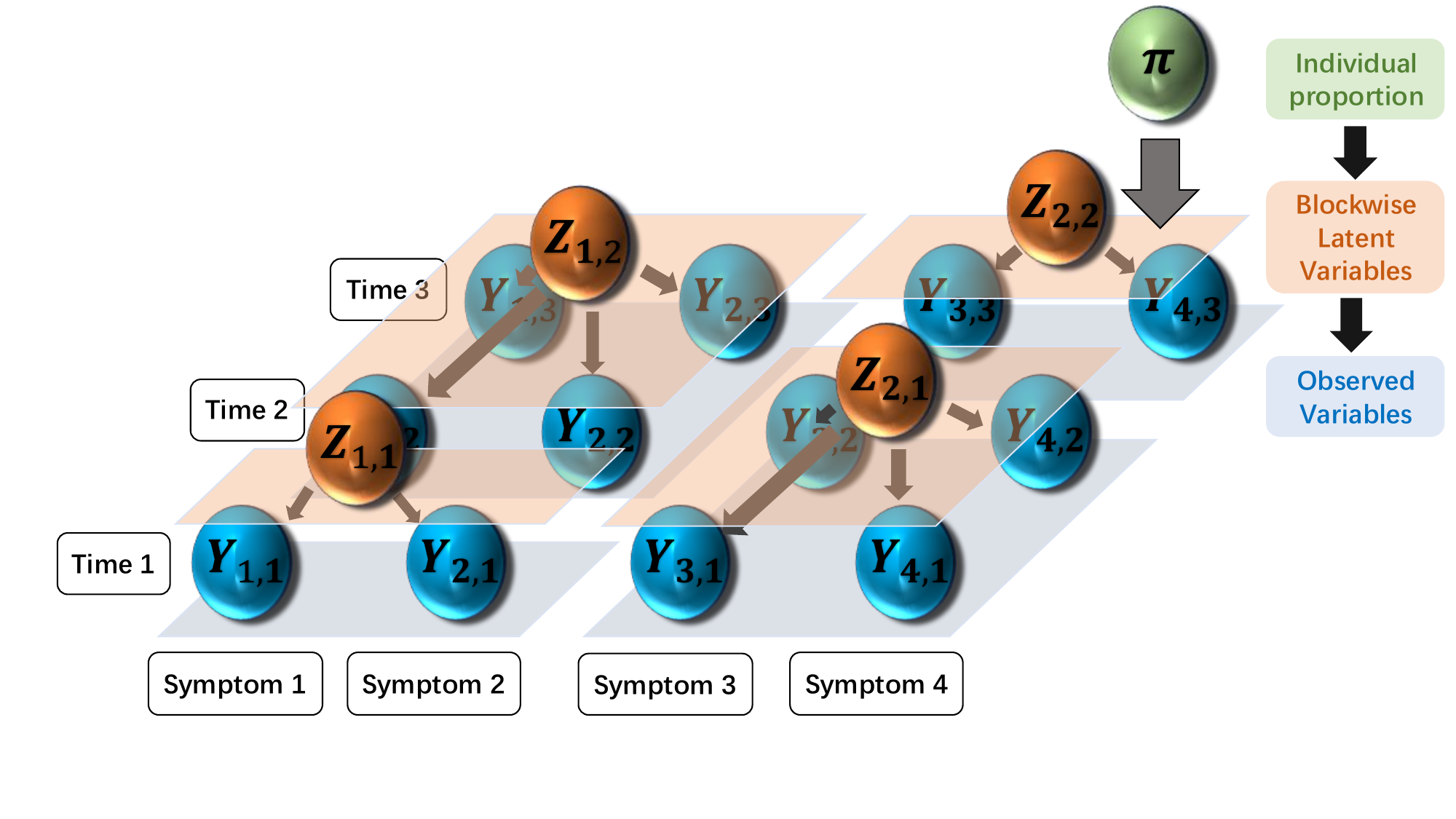} }}%
    \caption{Graphical representation of (a) conventional GoM and (b) BM$^3$. The model parameter $\bLambda$ is omitted for presentation simplicity.}%
    \label{fig:graphical}%
\end{figure}

We define $v_{r}$ as the cut-point for time periods such that ${b_{v_r}}=r$ and $b_{v_{r}+1}\neq r$, for $r = 1,\hdots,R-1$. By default, $v_{0} = 0$ and $v_{R} = T$. Since $y_{i,j,t}$ is categorical, the conditional distribution $f(y_{i,j,t}|\blambda_{j,k})$ is specified as $P(y_{i,j,t}|\blambda_{j,k}) = \prod_{c_j=1}^{d_j} \lambda_{j,c_j,k}^{I(y_{i,j,t}=c_j)}$ where $\blambda_{j,k} = (\lambda_{j,1,k},\ldots,\lambda_{j,d_j,k})^\top$. Then, the $BM^3$ involves the following generative process
\begin{enumerate}
    \item Draw group indicators $s_1, \hdots, s_p \sim $ Categorical$([G],\xi_1,\hdots,\xi_G)$.
    \item Draw period cut-points $v_{1},\hdots,v_{R-1} \sim $ Categorical $([T],\bone)$.
    \item For each subject $i$:
    \begin{itemize}
        \item[a.] Draw the proportion vector $\bpi_i \sim$ Dirichlet$(\balpha)$.
        \item[b.] Draw blockwise latent memberships $z_{i,1,1},\hdots,z_{i,G,R} \sim$ Categorical($[K],\bpi_i$).
        \item[c.] For each symptom $j$ and time point $t$, draw the observation $\{y_{i,j,t}\}_{s_{j}=g,b_{t}=r}|z_{i,g,r} = k \sim $ Categorical$([d_j],\lambda_{j,1,k},\hdots,\lambda_{j,d_j,k})$.
    \end{itemize}

\end{enumerate}
Denote $\bLambda = \{\blambda_{j,k},j=1,\hdots,p;k=1,\hdots,K\}$. Given the parameters $(\bb,\bs,\bLambda,\balpha)$, the probability mass function of $\{y_{i,1,1},\hdots,y_{i,p,T}\}$ can be written as
\begin{equation}\label{eq-pmf}
p(y_{i,1,1},\hdots,y_{i,p,T}|\bb,\bs,\bLambda,\balpha) = \int_{\Delta_{K-1}} \prod_{g=1}^G \prod_{r=1}^R \Big[\sum_{k=1}^K \pi_{i,k} \prod_{j:s_{j}=g} \prod_{t:b_{t}=r} \prod_{c_j=1}^{d_j} \lambda_{j,c_j,k}^{I(y_{i,j,t}=c_j)} \Big] dD_{\balpha}(\bpi_i).
\end{equation}
Figure \ref{fig:graphical} illustrates the graphical representations of the conventional GoM and the proposed BM$^3$. Notably, the distribution of mixed membership scores $\bpi_i$ is not restricted to a specific form. We choose Dirichlet distribution here for its popularity in mixed membership modeling for discrete data \citep{erosheva2007describing,wang2017variational,gu2023dimension}.

The BM$^3$, as defined in \eqref{eq-pmf}, subsumes several important statistical models as its special cases. For instance, if $G=R=1$, then the proposed BM$^3$ reduces to a latent class model, assuming that all $p\times T$ observations for subject $i$ belong to a single block. If $G=p$ and $R=T$, then BM$^3$ reduces to a conventional GoM, where each observation has its own latent membership. If one only fix $R=1$, then BM$^3$ reduces to the dimension-grouped mixed membership model proposed by \cite{gu2023dimension}, which imposes grouping structures on cross-sectional data. 
Compared to these models, our BM$^3$ offers enhanced capabilities in capturing the complex heterogeneity of PD longitudinally. Unlike conventional models that assume all observations are either independent (e.g., GoM) or fully dependent (e.g., latent class model) conditional on latent membership, our BM$^3$ allows for data-driven discovery of dependencies in PD-related markers. Specifically, symptoms that co-occur or consecutive time points exhibiting similar conditions are more likely to be grouped into the same block by our BM$^3$, sharing the same membership. Our BM$^3$ lets data speak for themselves and recover the potential symptom- and time-dependence structure of PD to a great extent. 

From a disease perspective, given that PD is an incredibly complex and multifaceted illness, the obtained symptom grouping $\bs$ and time grouping $\bb$ can comprehensively delineate the pathology of PD from multiple aspects, including syndromes (e.g., tremor, autonomic dysfunction) and periods (e.g., before 60yrs, after 60yrs).  At the individual level, instead of forcing each PD patient to belong to a single cluster, our BM$^3$ offers a nuanced and interpretable classification, where each patient may belong to multiple clusters, each representing the patients' condition in a specific disease domain during a particular time period. By estimating the blockwise latent membership $\bz_i=(z_{i,1,1},\hdots,z_{i,G,R})$, we are able to subdivide patients into smaller cohorts with common key features and similar progression subtypes, thus facilitating the development of targeted treatments.

\subsection{Heterogeneous periods, time-variant Parameters $\blambda_{j,k}$ and unbalanced $T_i$}
Several modeling assumptions in the BM$^3$ defined in Section \ref{sec:block} can be relaxed to address the complexities of PPMI data. First, the block structure of time points in the original model \eqref{eq-pmf-new} is assumed to be homogeneous among symptom groups. However, \cite{jack2010hypothetical} and \cite{wang2024multilayer} suggest that PD-related biomarkers (e.g., facial expression, tremor, speech etc.) may exhibit different rates of deterioration over time, implying a varying time-dependence structure across symptom groups. 
To capture this, a more flexible version of BM$^3$ introduces a symptom group-specific block structure $\bb_{g}$, allowing the partition of time to further depend on symptom group $g$ (see Figure \ref{fig:block}c). Second, the current model assumes that the generative process of observations $y_{i,j,t}$s is governed by the same parameter $\blambda_{j,k}$ for all time points within the same latent membership $z_{i,j,t}=k$. This assumption can be relaxed by further incorporating time-variant parameters $\blambda_{j,t,k}$ in the conditional distribution $P(y_{i,j,t}|\blambda_{j,t,k}) = \prod_{c_j=1}^{d_j} \lambda_{j,t,c_j,k}^{I(y_{i,j,t}=c_j)}$. By doing so, the temporal structure in $Y$ is influenced by both the underlying latent membership and time-dependent parameters. Finally, longitudinal studies like PPMI often face unbalanced designs, where subjects have differing numbers of visits due to dropout, loss to follow-up, or death. Denote $T_{max} = \text{max}_{i} T_i$ as the maximum number of visits among subjects. In this case, the time-period cut-points $v_{1},\hdots,v_{R-1}$ are drawn from  $\text{Categorical}([T_{max}],\bone)$. For each symptom group $g$, cut-points $v_{g,1},\hdots,v_{g,R-1}$ are similarly generated.

Given the aforementioned $\bb_{g}$, $\blambda_{j,t,k}$ and $T_i$, the probability mass function of $\{y_{i,1,1},\hdots,y_{i,p,T_i}\}$ under a BM$^3$ with symptom group-specific block structure, time-dependent parameters, and unbalanced number of time points can be written as 
\begin{equation}\label{eq-pmf-new}
p(y_{i,1,1},\hdots,y_{i,p,T_i}|\bb,\bs,\bLambda,\balpha) = \int_{\Delta_{K-1}} \prod_{g=1}^G \prod_{r=1}^R \Big[\sum_{k=1}^K \pi_{i,k} \prod_{j:s_{j}=g} \prod_{t:b_{g,t}=r} \prod_{c_j=1}^{d_j} \lambda_{j,t,c_j,k}^{I(y_{i,j,t}=c_j)} \Big] dD_{\balpha}(\bpi_i).\nonumber
\end{equation}
Since PD is progressive and irreversible, the obtained blockwise latent membership $Z_{i,g,r}$ are typically monotonic as $r$ increase from 1 to $R$. Therefore, the cut-points of the time periods can be naturally interpreted as inflection points in the progression of the disease, marking transitions from mild to severe stages. By estimating $v_{g,r}$ for $g=1,\hdots,G$, we can compare and establish the temporal ordering of deterioration across different symptom groups, therby identifying the symptoms or syndromes that can measure the earlier pathological changes. 

In addition, clinical trials for neurological disorders usually recruit cohorts with diverse genetic backgrounds and diagnostic statuses (e.g., healthy controls, prodromal, early PD) at baseline. For instance, \cite{lewis2005heterogeneity} and \cite{he2024joint} demonstrated that the age of onset of neurological disorders varies significantly across distinct observed or latent subpopulations. Therefore,  
if the population is composed of 
$C$ observed/latent subpopulations, a more general BM$^3$ can incorporate a symptom group- and subpopulation-specific block structure $\bb_{g,c}$. 
In this case, by evaluating $v_{g,c,r}$ for $c=1,\hdots,C$, our BM$^3$ allows for the detection of subpopulations that exhibit rapid PD progression. This is crucial in new experimental medicine trials, as these subpopulations are particularly suitable candidates for recruitment into trials of disease-modifying therapies aimed at slowing the progression of widespread neurodegeneration \citep{greenland2019clinical}.

\section{Model Identifiability}
The block structure is crucial for uncovering clinical heterogeneity and identifying PD subtypes, and the model parameters form the foundation for interpreting the subtypes.
It is thus critical to thoroughly investigate the identifiability issue to ensure that both the block structure and model parameters can be uniquely identified from the PPMI data, and guarantee the validity of the obtained syndromes and periods.
We start with blocks such as Figure \ref{fig:block}(b), and then show that the identifiability conclusions apply readily to the most flexible scenarios such as Figure \ref{fig:block}(d). 
Given a group indicator $\bs$ and a period indicator $\bb$, define a notation $\mathcal S_{g,r}\subseteq \{1,\ldots,pT\}$, which is a subset of the collection of items across time points satisfying that $(j,t)\in \mathcal S_{g,r}$ if and only if $s_j=g$ and $b_{t}=r$.
Consider the following marginal probability distribution for the observed vector $\bo y_{i,:,:}:=(y_{i,1,1},\hdots,y_{i,J,T})$: 
\begin{align}\label{eq:general}
&~p(y_{i,1,1},\hdots,y_{i,J,T}|\bb,\bs, \bLambda,\balpha) = \int_{\Delta_{K-1}} 
\prod_{g=1}^G \prod_{r=1}^R
\Big[\sum_{k=1}^K \pi_{i,k}
\prod_{(j,t)\in \mathcal S_{g,r}}
f(y_{i,j,t} \mid \blambda_{j,t,k})
\Big] dD_{\balpha}(\bpi_i),
\end{align} 
where $D_{\balpha}(\bpi_i)$ is the Dirichlet distribution with parameters $\balpha = (\alpha_1,\ldots,\alpha_K)$. For notational convenience, we denote the number of categories of item $j$ at time $t$ by $d_{j,t}$, and define $\bLambda_{j,t}$ to be a $d_{j,t}\times K$ conditional probability table containing $\blambda_{j,t,1},\ldots, \blambda_{j,t,K}$ as columns.
In the most general case in Figure \ref{fig:block}(d), the blocks of observations are not necessarily induced by a group indicator $\bs$ and a period indicator $\bb$, but rather just a general partition of the $pT$ observations. In this case, we can view $(g,r)$ as one single index that ranges in $\{1,\ldots,GR\}$, and the $GR$ sets $\{\mathcal S_{g,r}\}$ form a partition of $\{1,\ldots,pT\}$.
We define a $pT\times GR$ block indicator matrix $\V$ with binary entries indicating the block membership of each item at each time point. 
The rows of $\V$ are indexed by the items across all time points, and columns by all the $GR$ latent blocks. For $(j,t)\in[pT]$ and $(g,r)\in[GR]$, the entry $V_{(j,t), (g,r)} = \mathbbm{1}((j,t)\in \mathcal S_{g,r})$ is the binary indicator of whether item $j$ at time point $t$ belongs to the block $\mathcal S_{g,r}$.
The matrix $\V$ is mathematically equivalent to the notation of $\{\mathcal S_{g,r}: (g,r)\in[GR]\}$ but is more convenient for stating our identifiability conditions in the most general case. 

\begin{definition}[Strict Identifiability]
A BM$^3$ is said to be strictly identifiable if for any valid set of true parameters $(\V,\bLambda,\balpha)$ in the parameter space $\mathcal T$, the following equations hold if and only if the true parameters $(\V,\bLambda,\balpha)$ and alternative parameters $(\overline\V,\overline\bLambda,\overline\balpha)$ are identical up to the label swapping of the $K$ extreme latent profiles and that of the $GR$ blocks: 
		$\mathbb P(\bo y_{i,:,:} = \bo c\mid \V,\bLambda,\balpha)
		=
		\mathbb P(\bo y_{i,:,:} = \bo c\mid \overline\V, \overline\bLambda, \overline\balpha)$ for all
		$\bo c = (c_1,\ldots,c_{pT})\in \times_{t=1}^T \times_{j=1}^p [d_{j,t}]$.
\end{definition}

The above identifiability notion covers both (a) the discrete structure $\V$ about how the items across time are grouped into blocks and (b) the continuous parameters $\bLambda$ and $\balpha$. Once $\V$ is identified, all information about the blocks is identified. To see this, consider the block structures in Figure \ref{fig:block}(b), (c), and (d). Identifying $\V$ up to the column permutation (i.e., label swapping of the $GR$ blocks) is equivalent to identifying the different colors of the cells in the $p\times T$ table in each panel of Figure \ref{fig:block}. If there are well-defined time blocks and item blocks as in Figure \ref{fig:block}(b), then the original time-block vector $\bb = (1,1,1,2,2,2)$ and the item-block vector $\bs=(1,1,1,2,2,2)$ can be directly read off from the colors. 

Before stating the identifiability result, we introduce some notation.
Denote by $\bigotimes$ the Kronecker product of matrices and by $\bigodot$ the Khatri-Rao product \citep{koldabader2009}. 
For matrices $\mathbf A=(a_{i,j})\in\mathbb R^{m\times r}$, $\mathbf B=(b_{i,j})\in\mathbb R^{s\times t}$; 
and  $\mathbf C=(c_{i,j})=(\bo c_{:,1}\mid\cdots\mid\bo c_{:,k})\in\mathbb R^{n\times k}$,
$\mathbf D=(d_{i,j})=(\bo d_{:,1}\mid\cdots\mid\bo d_{:,k})\in\mathbb R^{\ell\times k}$, we have $\mathbf A\bigotimes \mathbf B \in\mathbb R^{ms\times rt}$ and $\mathbf C\bigodot \mathbf D \in\mathbb R^{n \ell\times k}$ with
\begin{align*}
	\mathbf A\bigotimes \mathbf B
	=
	\begin{pmatrix}
		a_{1,1}\mathbf B & \cdots & a_{1,r}\mathbf B\\
		\vdots & \vdots & \vdots \\
		a_{m,1}\mathbf B & \cdots & a_{m,r}\mathbf B
	\end{pmatrix},
	\qquad
	\mathbf C\bigodot \mathbf D
	=
	\begin{pmatrix}
		\bo c_{:,1}\bigotimes\bo d_{:,1}
		\mid \cdots \mid
		\bo c_{:,k}\bigotimes\bo d_{:,k}
	\end{pmatrix}.
\end{align*}
The Khatri-Rao product is the column-wise Kronecker product and will be useful in stating our following identifiability result.
\color{black}





\begin{theorem}[Strict Identifiability]
\label{thm-strid}
    Denote by $\mathcal{A}_{g,r} = \{(j,t)\in[p]\times[T]:\, (j,t)\in\mathcal S_{g,r}\}$ the set of variables that belong to block $(g,r)$.
    Suppose each $\mathcal{A}_{g,r}$ can be partitioned into three non-overlapping sets $\mathcal{A}_{g,r}^{(1)}$, $\mathcal{A}_{g,r}^{(2)}$, $\mathcal{A}_{g,r}^{(3)}$,
    and for each $m\in\{1,2,3\}$ the matrix $\bigodot_{(j,t)\in \mathcal{A}_{g,r}^{(m)}} \bo\Lambda_{j,t}$ has full column rank $K$.
    Also suppose for each $j\in[p]$ and $t\in[T]$, not all the column vectors of $\bo\Lambda_{j,t}$ are identical.
    Then the model parameters $\V$, $\bo\Lambda$, and $\balpha$ are strictly identifiable.
\end{theorem}

We emphasize that requiring the Khatri-Rao product $\bigodot_{(j,t)\in \mathcal{A}_{g,r}^{(m)}} \bo\Lambda_{j,t}$ (with $\prod_{(j,t)\in \mathcal{A}_{g,r}^{(m)}} d_{j,t}$ rows) to have full column rank $K$ as in Theorem \ref{thm-strid} is much weaker than requiring any individual matrix $\bo\Lambda_{j,t}$ (with $d_{j,t}$ rows) to have full column rank $K$. 
Next, we propose easier-to-check conditions for \emph{generic identifiability}, which is a slightly weaker notion than strict identifiability but often suffices for real data analysis purposes \citep{allman2009identifiability}.
Generic identifiability is proposed and popularized by \cite{allman2009identifiability}, roughly meaning that the parameters are identifiable almost everywhere in the parameter space, only except for a Lebesgue-measure zero subset.

\begin{definition}[Generic Identifiability]\label{def-genid}
    A BM$\,^3$ is said to be generically identifiable, if there is a subset of the parameter space $\mathcal N \subseteq \mathcal T$ with Lebesgue measure zero in $\mathcal T$ such that for any $(\bo\Lambda, \balpha)\in \mathcal T\setminus \mathcal N$ and a $\V$ matrix, the following holds if and only if $(\V, \bo\Lambda, \balpha)$ and the alternative $(\overline\V, \overline{\bo\Lambda}, \overline{\balpha})$ are identical up to permutations of the $K$ extreme latent profiles and that of the $GR$ variable groups:
		$\mathbb P(\bo y_{i,:,:} = \bo c\mid \V,\bLambda,\balpha)
		=
		\mathbb P(\bo y_{i,:,:} = \bo c\mid \overline\V, \overline\bLambda, \overline\balpha)$ for all
		$\bo c\in \times_{t=1}^T \times_{j=1}^p [d_{j,t}]$.
\end{definition}

\begin{theorem}[Generic Identifiability]
\label{thm-genid}
Still consider $\mathcal{A}_{g,r}$ defined in Theorem \ref{thm-strid} with the partition of it into three non-overlapping sets $\mathcal{A}_{g,r} = \cup_{m=1}^3 \mathcal{A}_{g,r}^{(m)}$. Parameters $\bo\Lambda$, $\V$, and $\bo\Phi$ are generically identifiable if the following holds for each $(g,r)\in[G]\times[R]$ and $m=1,2,3$:
    \begin{align}\label{eq-djk}
        \prod_{(j,t)\in\mathcal{A}_{g,r}^{(m)}} d_{j,t} \geq K,
    \end{align}
\end{theorem}
Comparing Theorem \ref{thm-genid} to Theorem \ref{thm-strid}, the full-rank requirements on certain Khatri-Rao products of the unknown parameters are lifted and replaced by a simple requirement based on comparing some numbers of the response patterns in the blocks with the number of extreme latent profiles. Intuitively, Condition \eqref{eq-djk} in Theorem \ref{thm-genid} is easy to satisfy as long as there are enough items (i.e., symptoms) or enough categories of items. As an example, consider Figure \ref{fig:block}(b) with $G=2$ and $R=2$, and suppose each item response has $d_{j,t}=3$ categories. Then the four sets $\mathcal{A}_{1,1}$, $\mathcal{A}_{1,2}$, $\mathcal{A}_{2,1}$, and $\mathcal{A}_{2,2}$ each contains 9 items, and if we let each $\mathcal{A}_{g,r}^{(m)}$ contain 3 items for $m=1,2,3$, then $\prod_{(j,t)\in\mathcal{A}_{g,r}^{(1)}} d_{j,t} =  \prod_{(j,t)\in\mathcal{A}_{g,r}^{(2)}} d_{j,t} =  \prod_{(j,t)\in\mathcal{A}_{g,r}^{(3)}} d_{j,t} = 3^3 = 27$. So condition \eqref{eq-djk} will be satisfied as long as $27\geq K$. In other words, in this case of Figure \ref{fig:block}(b), the proposed BM$^3$ is generically identifiable even if there are as many as $K=27$ extreme latent profiles, which corresponds to a very expressive and flexible model.

\begin{theorem}[Posterior Consistency]\label{thm-posterior}
Suppose the conditions for strict identifiability are satisfied and the prior distributions have full support around the true parameter values. Then the posterior distributions for the model parameters will concentrate around the true parameter values as sample size $n$ goes to infinity.
\end{theorem}

The posterior consistency result ensures valid Bayesian posterior inference on both the continuous model parameters and the discrete block structure of items in the model.

\section{Bayesian Inference}
We adopt a Bayesian approach with a MCMC algorithm for posterior inference. We first specify the prior distributions for all unknown parameters in BM$^3$ with symptom group-specific periods. 
The details of Bayesian inference for BM$^3$ with symptom group- and subpopulation-specific periods are provided in Supplementary Material. 
Let $\bch = \{(v_{1},\hdots,v_{R-1}),1\leq v_1<\hdots< v_{R-1}\leq T-1\}$ denotes the set containing in total $C_{T,R-1}$ possible vectors of cut-off points for time periods, where $C_{T,R-1}$ represents the combination number. For the symptom group indicator $s_1,\hdots,s_p$ and cut-off points of symptom group-specific time period $(v_{1},\hdots,v_{R-1})$, we assign the following categorical distribution as their priors:
\begin{align*}
s_1,\hdots,s_p &\overset{\text{iid}}{\sim} \text{Categorical}([G],\xi_1,\hdots,\xi_G)\\
(v_{g,1},\hdots,v_{g,R-1}) &\sim \text{Categorical}([C_{T,R-1}],\kappa_{g,1},\hdots,\kappa_{g,C_{T,R-1}})
\end{align*}
where Categorical($[G],\xi_1,\hdots,\xi_G$) denotes a categorical distribution over $G$ categories with proportions $\xi_1,\hdots,\xi_G$. For $(\xi_1,\hdots,\xi_G)$, $(\kappa_{g,1},\hdots,\kappa_{g,C_{T,R-1}})$ and $\blambda_{j,k}$, we assign uniform priors over the  probability simplex. For the parameters $\balpha$ in Dirichlet distributions, let $\alpha_0 = \sum_{k=1}^K \alpha_k$ and $\bfeta = (\alpha_1/\alpha_0,\hdots,\alpha_K/\alpha_0)$ denote the Dirichlet parameters. We assign a gamma prior $\text{Gamma}(a_{\alpha},b_{\alpha})$ and a uniform over $K-1$ probability simplex for $\alpha$ and $\bfeta$, respectively. For notational simplicity, define indicators $z_{i,g,r,k}=I(z_{i,g,r}=k)$ and $y_{i,j,t,c} = I(y_{i,j,t}=c)$. We adopt a Metropolis-within-Gibbs sampler that cycles through the following steps.\\
1. For each $j\in[p]$ and $k\in[K]$, sample the conditional probabilities $\blambda_{j,k}$s from their posterior distributions:
\begin{align}
    \{\blambda_{j,k}|-\}_{s_j=g} &\sim \text{Dirichlet}\Big(1+\sum_{i=1}^n\sum_{r=1}^R\sum_{t=v_{g,r-1}+1}^{v_{g,r}} z_{i,g,r,k}y_{i,j,t,1},\hdots,1+\sum_{i=1}^n\sum_{r=1}^R\sum_{t=v_{g,r-1}+1}^{v_{g,r}} z_{i,g,r,k}y_{i,j,t,d_j} \Big).\nonumber
\end{align}
2. For each subject $i\in[n]$, group $g\in[G]$ and each period $r\in[R]$, sample the mixed membership score $\bpi_i$ and latent allocation variable $z_{i,g,r}$ from their posterior distributions:
\begin{align}
        \bpi_i|-&\sim \text{Dirichlet}\Big(\alpha_1+\sum_{g=1}^G\sum_{r=1}^R z_{i,g,r,1},\hdots,\alpha_K+\sum_{g=1}^G\sum_{r=1}^R z_{i,g,r,K} \Big),\nonumber\\
    P(z_{i,g,r}=k|-)&= \frac{\pi_{i,k} \prod_{j:s_j=g}\prod_{t=v_{g,r-1}+1}^{v_{g,r}}\prod_{c=1}^{d_j} \lambda_{j,c,k}^{y_{i,j,t,c}} }{\sum_{k^{\prime}=1}^K \pi_{i,k^{\prime}} \prod_{j:s_j=g}\prod_{t=v_{g,r-1}+1}^{v_{g,r}}\prod_{c=1}^{d_j} \lambda_{j,c,k^{\prime}}^{y_{i,j,t,c}}}, \ \ \ k\in [K].\nonumber
\end{align}
3. For each group $g\in[G]$ and period $r\in[R]$, sample the block structure $(s_1,\hdots,s_p)$ and $(v_{1},\hdots,v_{R-1})$. Denote $\bh_m=(h_{m,1},\hdots,h_{m,R-1})$ as the $m$-th element in $\bch$, $m=1,\hdots,C_{T,R-1}$. The posterior distributions of $(s_1,\hdots,s_p)$ and $\bv_g=(v_{g,1},\hdots,v_{g,R-1})$ are
\begin{align}   
    P(s_j=g|-)&= \frac{\xi_{g} \prod_{i=1}^n \prod_{r=1}^R \prod_{t=v_{g,r-1}+1}^{v_{g,r}} \lambda_{j,y_{i,j,t},z_{i,g,r}} }{\sum_{g^{\prime}=1}^G \xi_{g^{\prime}} \prod_{i=1}^n \prod_{r=1}^R \prod_{t=v_{g,r-1}+1}^{v_{g,r}} \lambda_{j,y_{i,j,t},z_{i,g^{\prime},r}}}, 
    \nonumber\\
    (v_{g,1}=h_{m,1},\hdots,v_{g,R-1}=h_{m,R-1}|-) 
    &= \frac{\kappa_{m} \prod_{i=1}^n 
    \prod_{j:s_{j}=g} \prod_{r=1}^R \prod_{t=h_{m,r-1}+1}^{h_{m,r}} \lambda_{j,y_{i,j,t},z_{i,g,r}} }{\sum_{m^{\prime}=1}^{C_{T,R-1}} \kappa_{m^{\prime}} \prod_{i=1}^n
    \prod_{j:s_{j}=g} \prod_{r=1}^R \prod_{t=h_{m^\prime,r-1}+1}^{h_{m^\prime,r}} \lambda_{j,y_{i,j,t},z_{i,g,r}}}.\nonumber
    \end{align}
    The posterior distributions of $(\xi_1,\hdots,\xi_G)$ and $(\kappa_{1},\hdots,\kappa_{C_{T,R-1}})$ are
    \begin{align}
        (\xi_1,\hdots,\xi_G|-)&\sim \text{Dirichlet}\Big(1+\sum_{j=1}^p I(s_j=1),\hdots,1+\sum_{j=1}^p I(s_j=G) \Big),\nonumber\\
    (\kappa_{g,1},\hdots,\kappa_{g,C_{T,R-1}}|-) &\propto \text{Dirichlet}\Big(1+ I(\bv_g=\bh_1),\hdots,1+I(\bv_g=\bh_{C_{T,R-1}}) \Big).\nonumber
\end{align}
4. Sample the Dirichlet parameters $\balpha= (\alpha_1,\hdots,\alpha_K)$ via a Metropolis-Hastings algorithm. The full conditional distribution of $\balpha$ is
\begin{align}
    (\balpha|-)\propto \nonumber \alpha_0^{a_\alpha - 1}\exp(-\alpha_0 b_{\alpha})\times \Big[\frac{\Gamma(\alpha_0)}{\prod_{k=1}^K \Gamma(\alpha_k)}\Big]^n,
\end{align}
which does not have an analytic form. We adopt a Metropolis-Hastings algorithm as follows.\\
(4a) Sample the proposal $\alpha^* = (\alpha_{1}^*,\hdots,\alpha_{K}^*)$ from log-normal distribution $\alpha_{k}^* \overset{\text{iid}}{\sim} \text{lognormal}(\log \alpha_k,\sigma_\alpha^2) $.\\
(4b) Accept $\alpha^*$ with probability
\begin{align}
    r = \min\Big\{1,e^{-a_\alpha(\alpha_0^*-\alpha_0)}\frac{\alpha_0^*}{\alpha_0}^{b_\alpha-1}\big(\prod_{k=1}^K \frac{\alpha_k^*}{\alpha_k}\big)\Big[\frac{\Gamma(\alpha_0^*)}{\Gamma(\alpha_0)}\prod_{k=1}^K \frac{\Gamma(\alpha_k^*)}{\Gamma(\alpha_k)} \Big]^n \prod_{k=1}^K \big(\prod_{i=1}^n g_{ik}\big)^{\alpha_k^*-\alpha_k} \Big\},\nonumber
\end{align}
where $\alpha_0^* = \sum_{k=1}^K \alpha_k^*$, and $\sigma_\alpha^2$ is chosen to achieve an acceptance rate around $60\%$.

Model selection is a critical issue in GoMs,  as the number of extreme profiles $K$ is usually unknown in real-world applications and must be carefully determined. This challenge becomes even more pronounced in the proposed BM$^3$, where the number of blocks $G\times R$ is also unknown. Broadly speaking, two primary approaches have been developed for order selection in GoMs and related latent class models. One approach involves regularization-based techniques. For instance, \cite{chen2009order} proposed a nonsmooth penalty to avoid near-zero mixing proportions and similar components, thereby selecting the optimal number of mixture components for finite mixture models. 
However, in the proposed BM$^3$, the additional complexity introduced by the unknown block structure makes applying these penalization techniques challenging. The second approach relies on criterion-based methods to determine the optimal model. For example, \cite{erosheva2007describing} employed Akaike’s information criterion (AIC) for MCMC samples \citep{raftery2006estimating} to select the number of extreme profiles in GoMs for multivariate binary data. 
Recently, \cite{gu2023dimension} employed the Widely Applicable Information Criterion \cite[WAIC,][]{watanabe2010asymptotic} to determine the number of extreme profiles for dimension-grouped mixed membership models. Notably, \cite{watanabe2010asymptotic} highlighted that WAIC is particularly well-suited for models with hierarchical and mixture structures, making it an appropriate choice for selecting both the block structure and the number of extreme profiles in BM$^3$. 

Let $\btheta^{(l)}, l=1,\hdots,L$ represent the posterior samples, where $L$ is the total number of samples. For each $i \in [n]$ and $l\in[L]$, let
\begin{align}
    p(\by_i,\btheta^{(l)}) = \prod_{g=1}^G \prod_{r=1}^R \Big[\sum_{k=1}^K \pi_{i,k} \prod_{j:s_{j}=g} \prod_{t:b_{t}=r} \prod_{c_j=1}^{d_j} \lambda_{j,c_j,k}^{I(y_{i,j,t}=c_j)} \Big].\nonumber
\end{align}
We adopt the version of WAIC recommended by \cite{gelman2014understanding}, given by:
\begin{align}
    \text{WAIC} = -2 \sum_{i=1}^n \log\Big(\frac{1}{L}\sum_{l=1}^Lp(\by_i,\btheta^{(l)})\Big)+2\sum_{i=1}^n\text{var}_{l=1}^L(\log p(\by_i,\btheta^{(l)})),\nonumber
\end{align}
where the first term is log pointwise predictive density and the second term measures the model complexity by the sample variance. Models with smaller WAIC values are preferred. 

\section{PPMI Data Analysis}
\label{sec:realdata}
\subsection{PPMI dataset and preprocessing}
The PPMI is a longitudinal study designed to collect a comprehensive range of data on PD, including clinical features, imaging outcomes, biologic and genetic markers, and digital outcomes across all stages of the disease, from prodromal to moderate PD \citep{marek2011parkinson}. The overarching objective of PPMI is to identify biomarkers of disease progression, ultimately facilitating the development of effective interventions and targeted treatments for PD-related disability. One of the main modalities within PPMI is the assessment of clinical symptoms, which capture both motor disorders, such as tremor and rigidity, and nonmotor features, such as constipation and sleeping disturbance. 
A complication of analyzing the PPMI dataset is the substantial heterogeneity in disease progression, as measured by clinical symptoms.
At the individual level, patients may experience varying trajectories of disease progression—some progressing through all stages, from unilateral symptoms to requiring assistance for standing or walking, while others may skip stages.
At the symptom level, some symptoms occur independently,  while others co-occur with related symptoms. 
At the time level, serial dependence often exists among repeated measurements of clinical symptoms. The BM$^3$ offers several advantages for addressing these complexities:\\
(i) Between-subject Heterogeneity: BM$^3$ accounts for individual variability by allowing subjects to belong to distinct disease profiles.\\
(ii) Between-Symptom Heterogeneity: The model partitions co-occurring symptoms into symptom groups, assigning the same latent membership to observations within a group.\\
(iii) Between-Time Heterogeneity: BM$^3$ combines consecutive time points into periods with shared latent memberships and allows different symptom groups and subpopulations to have distinct temporal structures.

For this analysis, we focus on the MDS-UPDRS assessments, which consist of four parts with a total of 50 items. Each item has five response options, ranging from normal to severe, reflecting various aspects of PD symptomatology such as fatigue, dystonia, tremor, and bradykinesia. Since very few responses fall into the most severe category, we combine it with the moderate category.
Thus, each item has $d_j=4$ categories: 1 (normal), 2 (slight), 3 (mild), 4 (moderate to severe).  
Of the 50 items, 21 display little variability across subjects, so we focus on the remaining $p=29$ items that show more variability for the subsequent analysis. Our dataset includes 1,531 subjects with diverse genetic backgrounds and diagnostic statuses, as determined by neurologists at baseline. Each subject has at least one visit, with up to 18 follow-up visits. The ages of the subjects range from 26.4 to 92.6 years.
This study aims to use the proposed BM$^3$ to systematically discover the aforementioned between-patient, between-symptom, and between-time clinical heterogeneity within PD course and identify important PD subtypes.

\subsection{Fitting BM$^3$ on PPMI dataset}

We applied the proposed BM$^3$ with a symptom group- and subpopulation-specific block structure to the PPMI data. 
The hyperparameters were specified as $a_\alpha = 2, b_\alpha = 1$ and $\sigma_\alpha = 0.02$. 
Since our primary interest lies in identifying inflection points that indicate the transition from normal to severe stages of PD, we fixed the number of periods at $R=2$ and focused on inferring the cut-points for the two time periods.
We considered combinations of $G=\{2,3,4,5\}$, $K=\{2,3,4\}$ and $C=\{1,2,3\}$, leading to 36 competing models, and used the WAIC to select the optimal model. The MCMC algorithm was run for 10,000 iterations, with the first 5,000 iterations discarded as burn-in. The model with $G=4, K=3$ and $C = 2$ achieves the smallest WAIC value and was selected as the final model. 

\begin{figure}
    \centering
    \subfloat[\centering Symptom grouping structure]
    {{\includegraphics[width=8cm]{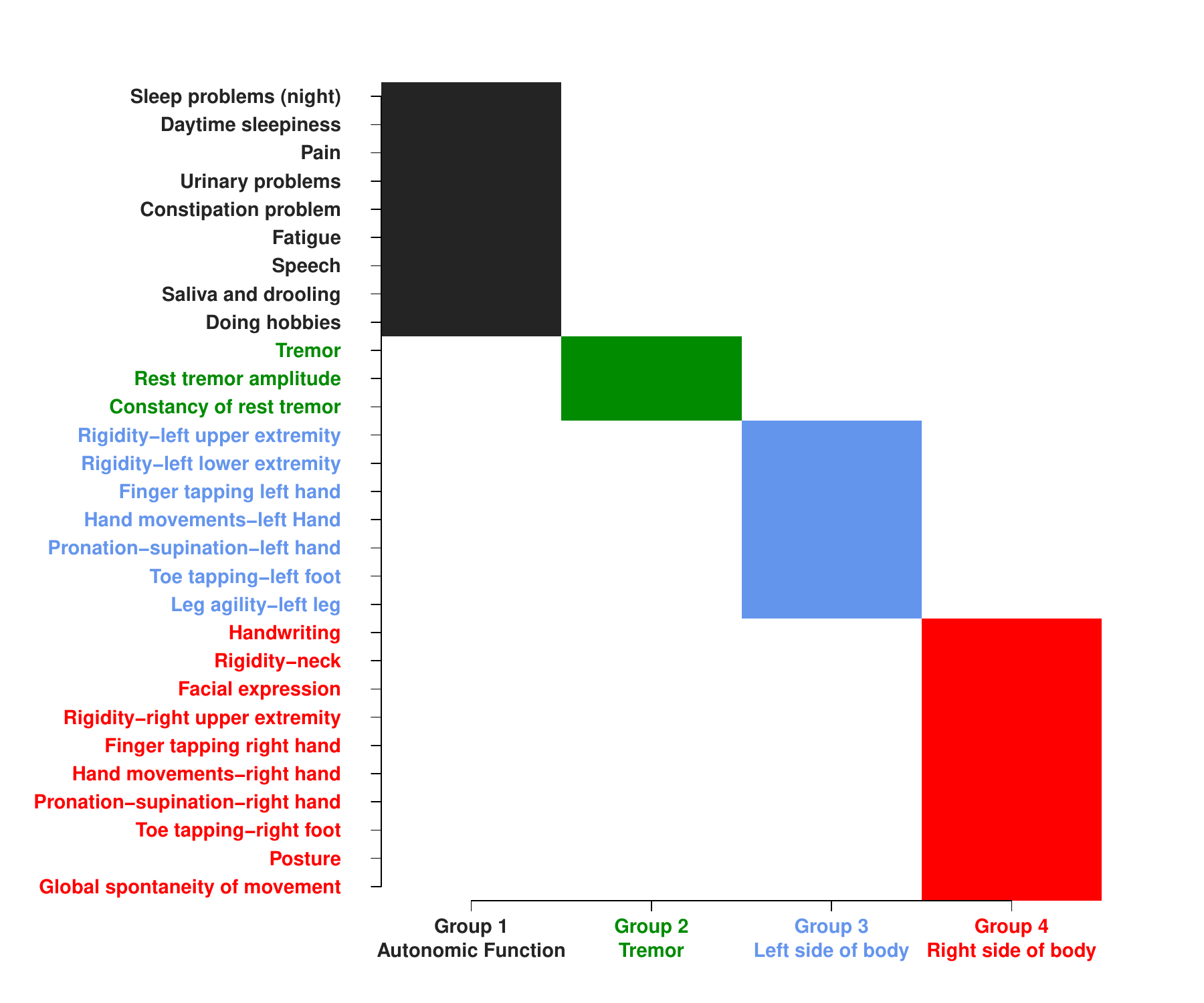} }}
    \subfloat[\centering Cut-points of the time periods]
    {{\includegraphics[width=8cm]{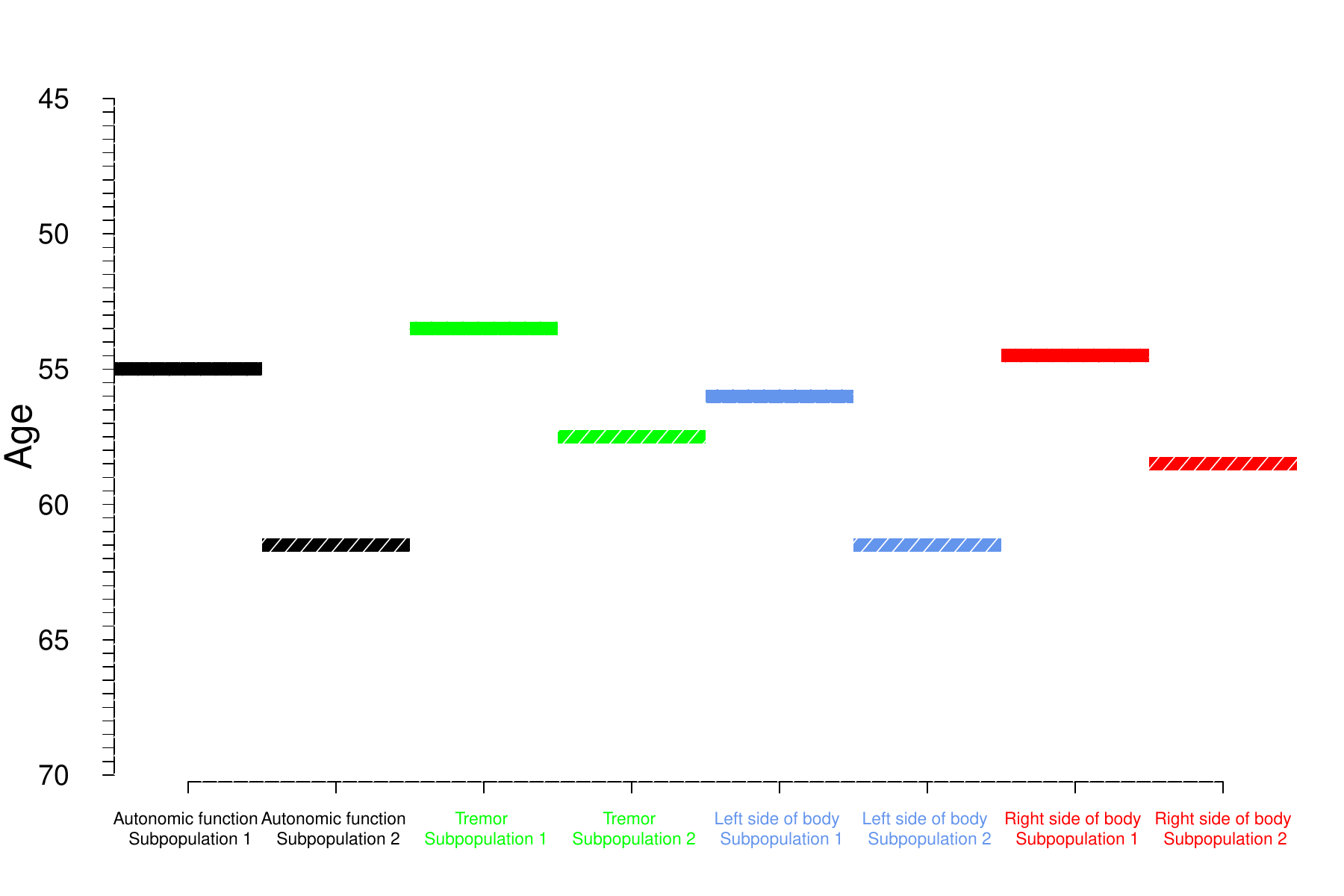} }}
    \caption{Estimated block structure in PPMI dataset analysis: (a) Symptom grouping structure of MDS-UPDRS assessment; (b) cut-points of the time periods for two subpopulations.}%
    \label{pic:group_mat}%
\end{figure}

From the fitted model, we calculated the posterior mode of $\bs$ to estimate the symptom grouping structure. Figure \ref{pic:group_mat} presents the symptom grouping structure of the 29 MDS-UPDRS items, which aligns well with clinical insights. Based on the clinical characteristics of these symptoms, the $G=4$ symptom groups can be interpreted as autonomic function, tremor, left-side motor functions, and right-side motor functions. Specifically, Group 1 consists of nine autonomic symptoms (e.g., saliva and drooling, sleep disturbances, and constipation), while Group 2 consists of three tremor-related assessments (e.g., tremor, rest tremor amplitude, and constancy of rest tremor). The remaining 17 movement disorder items are divided into Groups 3 and 4 based on whether they affect the left or right side of the body. For example, finger tapping on the left and right hands is assigned to Groups 3 and 4, respectively. These findings not only confirm the presence of between-symptom heterogeneity in PD but also suggest that neurologists should assess PD patients by examining these four clinical domains separately, rather than aggregating the scores of all 29 symptoms.


For the time period blocks, we summarized the posterior mode of $\bb_{g,c}$ in $G=4$ symptom groups and $C=2$ subpopulations. For the first subpopulation, 
The estimates $\bb_{g,1} = \bone_{18}$ for $g=1,\hdots,4$ indicate that all visits belong to the same time period (i.e., latent membership is homogeneous across time). This result suggests that patients in the first subpopulation exhibit stable symptoms across all four clinical domains throughout the disease course. In contrast, for the second subpopulation, we obtained the estimates of time-block structure $\bb_{1,1} = (\bone_{12}^\top,2*\bone_{6}^\top)$, $\bb_{2,1} = (\bone_{15}^\top,2*\bone_{3}^\top)$, $\bb_{3,1} = (\bone_{13}^\top,2*\bone_{5}^\top)$ and $\bb_{4,1} = (\bone_{17}^\top,2*\bone_{1}^\top)$.  This suggests that the inflection point for autonomic dysfunction occurs earlier than for tremor and movement complications, a finding that contrasts with existing literature  \citep{wang2024multilayer}. A possible explanation is that age is a major risk factor for PD progression, and since participants entered the PPMI study at different ages, partitioning observations by follow-up visits may not fully capture the true progression of the disease.

To account for the potential age effect, we reorganized the longitudinal data by age at each visit instead of using visit codes. Specifically, we partitioned age from the minimum (26.5 years) to the maximum (96 years) into $T=136$ half-year intervals. Each clinical assessment within a given interval was treated as a longitudinal observation at the corresponding time point, ensuring that patients retained the same number of observations as in the original analysis. We then refitted the BM$^3$ model with $G=4, K=3$ and $C=2$ on the rearranged data. The variable grouping structure from this analysis was identical to that in Figure \ref{pic:group_mat}(a), reinforcing the significant between-symptom heterogeneity in PD. However, the time period estimates differed from the previous analysis and provided more interpretable results. Figure \ref{pic:group_mat}(b) presents the estimated time periods for the two subpopulations. The cut-points for the second subpopulation consistently occurred later than those for the first subpopulation, indicating that patients in the second subpopulation experience slower disease progression. Additionally, the cut-points varied across the four symptom groups. Notably, the Tremor group exhibited the earliest cut-point, while the Autonomic Function group had the latest (or the second latest for subpopulation 1), suggesting that PD deterioration typically begins with tremor, followed by movement symptoms on right side of the body, and eventually affects autonomic functions and left-side motor function.

\begin{figure}[h!]
            \centering
            \includegraphics[width=0.7\textwidth]{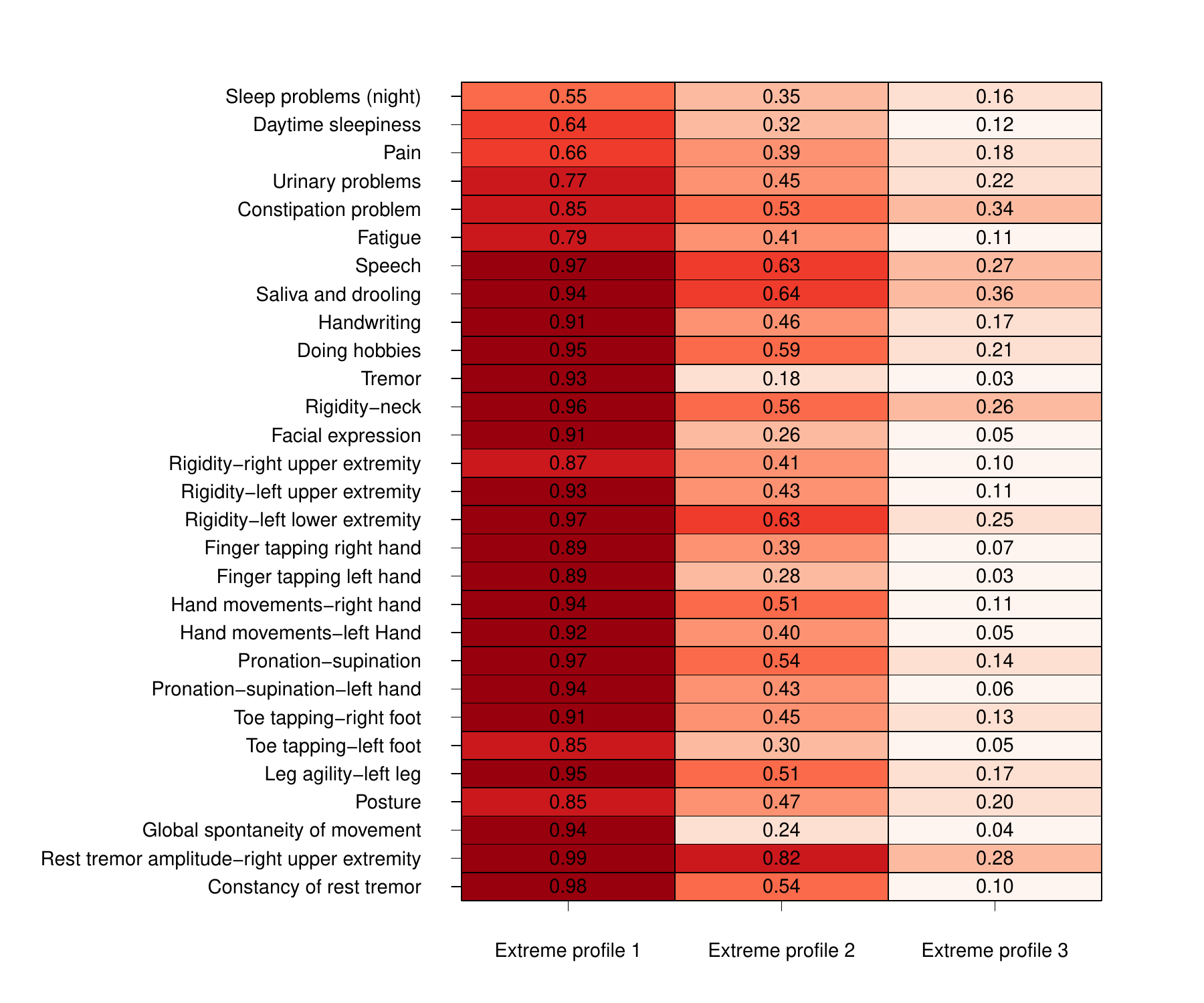}
            \caption{Estimated parameters $\lambda_{:,1,:}$ for the PPMI data. Each column represents one extreme latent profile. Entries are conditional probabilities of having a normal symptom (1 = normal) to each item given that extreme latent profile.} 
\label{pic:lambda}
\end{figure}

Finally, for the event probabilities, we present the posterior mean of $\blambda_{:,1,:}$ in Figure \ref{pic:lambda}. For each symptom $j$ and each extreme profile $k$, $\blambda_{j,1,k}$ represents the probability of having normal function on this symptom, conditional on belonging to the $k$-th extreme profile. The $K = 3$ extreme profiles are well separated and can be interpreted as normal function, moderate dysfunction and severe dysfunction, respectively. For instance, from extreme profile 1 to 3, the probability of having normal function in toe tapping decreases significantly. Notably, we observe a significant decline in $\blambda_{:,1,:}$ for two specific symptoms--tremor and Global spontaneity of movement--from extreme profile 1 to 2, implying substantial deterioration in these two symptoms as  the disease profile becomes more severe. 

\subsection{Latent blockwise membership representation}
\begin{figure}[h!]
            \centering
            \includegraphics[height=0.5\textheight]{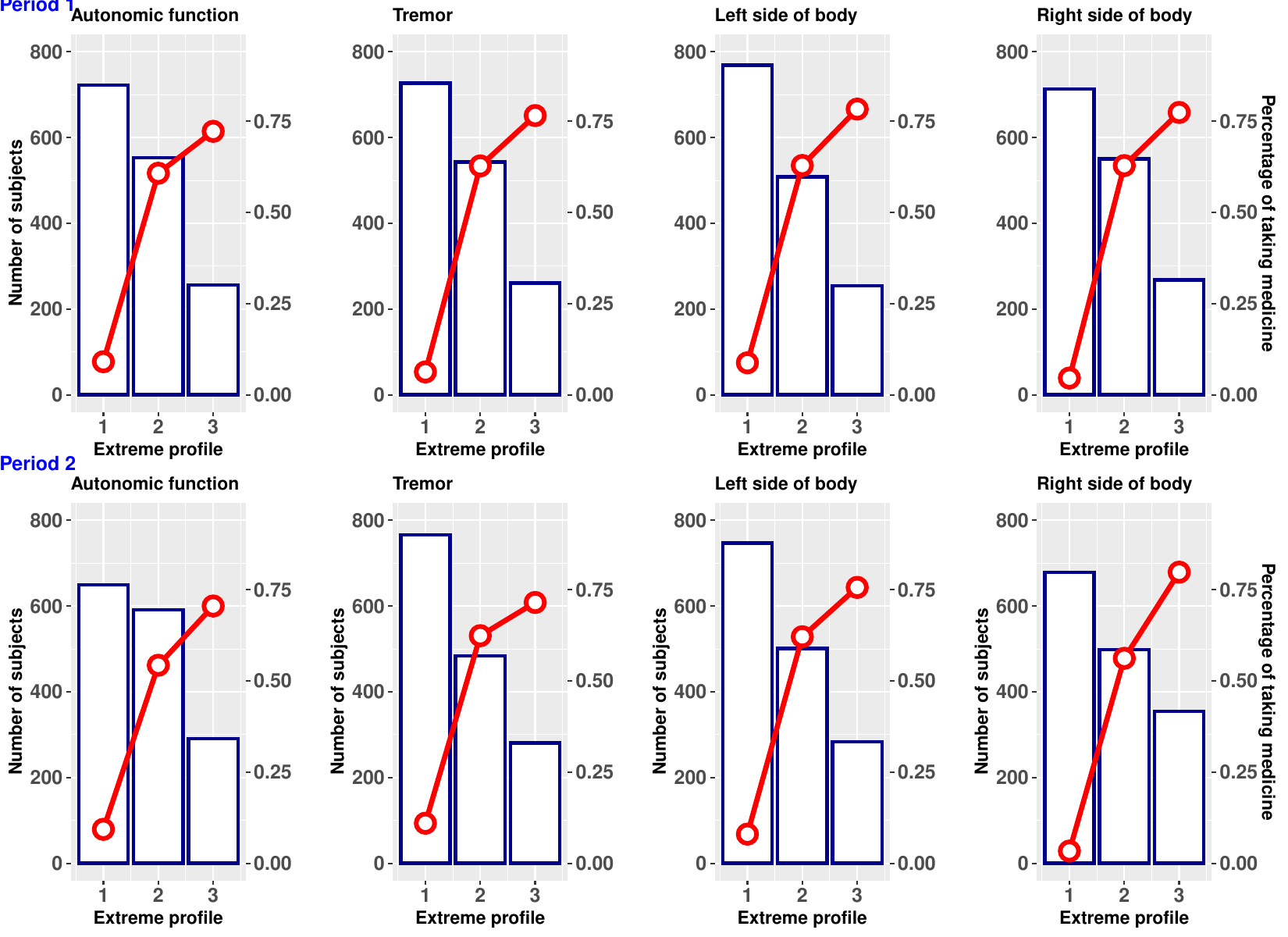}
            \caption{Bar plots: number of subjects belonging to each of three extreme profiles across four symptom groups and two periods. Red line charts: proportion of taking levodopa medication in each extreme profile.} 
\label{pic:Z_1}
\end{figure}

The blockwise latent membership $\{Z_{i,g,r}\}$ contains valuable information about subjects' statuses and disease progression learned from the PPMI dataset. We first calculate the posterior mode of each $Z_{i,g,r}$. Figure \ref{pic:Z_1} displays bar plots of the estimated $Z_{i,g,r}$ for each symptom group and time period. In general, most subjects belong to the first extreme profile, while the third profile has the fewest subjects across symptom groups and periods. Notably, the number of subjects assigned to the third extreme profile, which represents severe dysfunction, increases from period 1 to period 2. This indicates a gradual deterioration in patients’ disease status over time.
In addition, Figure S.1 in Supplementary Material presents sankey diagrams that visualize the changes in latent membership $Z_{g,r}$ across periods for each syndrome $g$. The most frequent transitions are from ``Normal" to ``Moderate" and from ``Moderate" to ``Severe", which aligns with the irreversible neurodegeneration seen in Parkinson’s disease (PD). This pattern supports our interpretation of the time period cut-offs as inflection points for PD-related syndromes. To further demonstrate the reliability and validity of the estimated blockwise latent memberships, we examined clinical variables that have not been used in the model fitting. One clinically important variable is whether the subject initiated levodopa medication, a dopamine replacement agent for the treatment of Parkinson disease, during the study follow-up. We calculated the proportion of subjects in each extreme profile who began levodopa medication. Figure \ref{pic:Z_1} shows that the proportion of levodopa use increases progressively from extreme profile 1 to 3 across all symptom groups and time periods. This finding suggests that the blockwise latent variables obtained by the proposed BM$^3$ contain substantial amount of information of disease course.

\begin{figure}[h!]
            \centering
            \includegraphics[width=\textwidth]
            {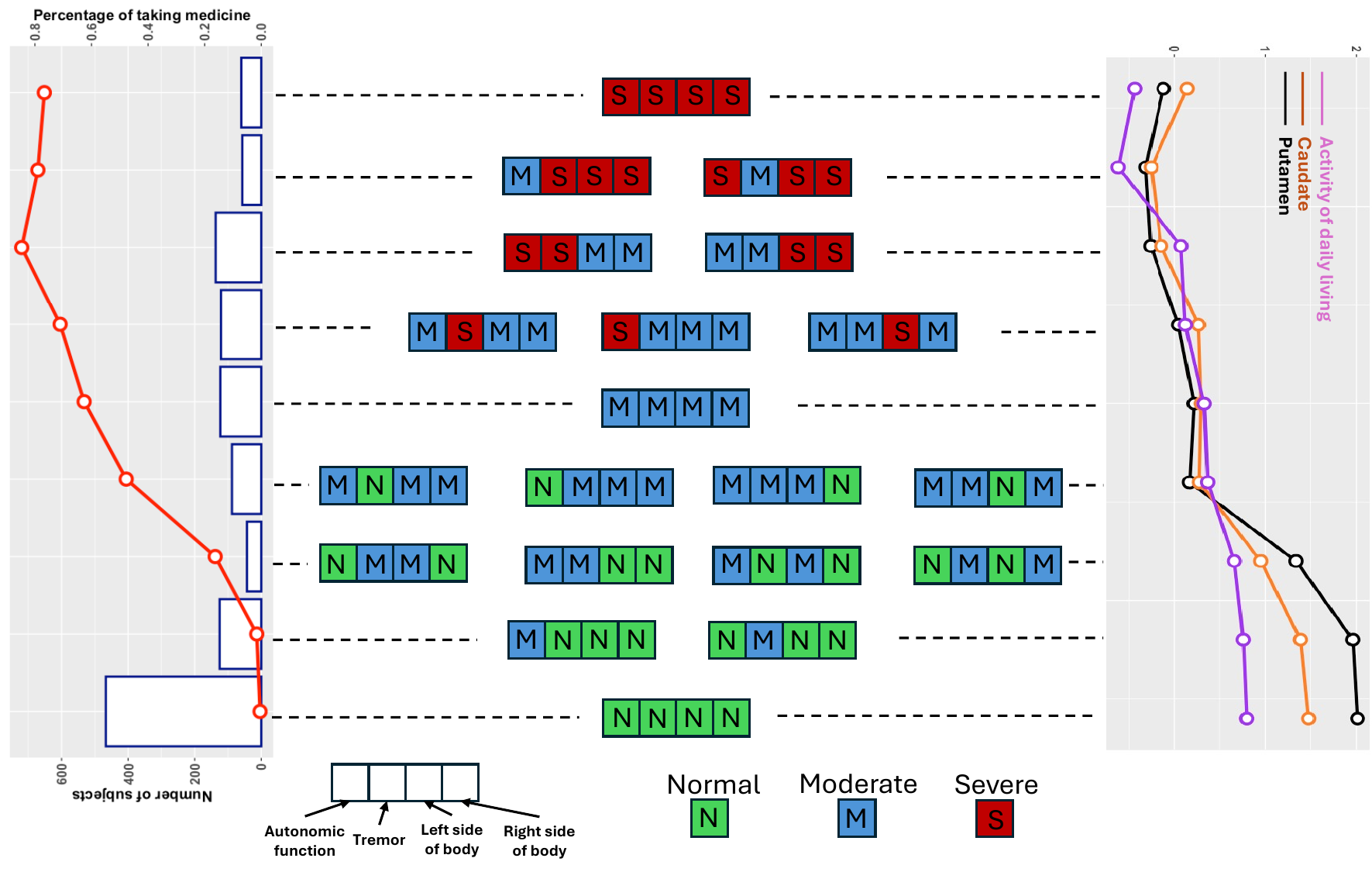}
            \caption{The 20 most frequent PD subtypes identified from the PPMI study, each corresponding to a combination of latent memberships in four clinical domains.
            Extreme profile 1 to 3 are denoted as ``N", ``M", and ``S", respectively. 
            Left bar plot: the number of subjects in each subtype. Left line chart: proportion of taking levodopa medication in each subtype. Right line charts: averaging measures of external biomarkers for subjects in certain subtypes.} 
\label{fig:factor2}
\end{figure}

Beyond single latent membership in each clinical domain, we combined the latent memberships from all four symptom groups together and thus partition subjects into $K^G = 3^4 = 81$ potential subtypes. Figure \ref{fig:factor2} presents the 20 most frequent disease subtypes. For instance, subjects with $(Z_{1,1},Z_{2,1},Z_{3,1},Z_{4,1}) = (2,3,2,2)$ represent a typical PD subtype characterized by moderate dysfunction in autonomic function and both sides of the body, but severe dysfunction in tremor (denoted as ``M$|$S$|$M$|$M" in Figure \ref{fig:factor2}). Among the disease subtypes, the most severe subtype, ``S$|$S$|$S$|$S", represents severe dysfunctions across all four clinical domains while the healthiest subtype, ``N$|$N$|$N$|$N", corresponds to subjects with normal function in all domains. The remaining 18 subtypes are presented in a partially ordered structure (e.g., ``M$|$S$|$M$|$M" is comparable to ``S$|$M$|$M$|$M", but more severe than ``M$|$M$|$M$|$M") from top to bottom in Figure \ref{fig:factor2}. These disease subtypes have significant therapeutic implications. First, identifying these subtypes is essential for developing more targeted and effective treatments. Second, these subtypes can guide the selection of the most appropriate patients for clinical trials.

To further validate the subtypes, we again used external variables. We observed that the proportion of subjects taking levodopa medication gradually increase as the severity of the disease subtypes escalates (see left line chart in Figure \ref{fig:factor2}). Additionally, we explored whether the subtypes obtained from the BM$^3$ are consistent with other behavioural measures or even neuroimaging biomarkers. Intuitively, if the estimated combined latent membership characterize the overall disease status of patients, then these subtypes should also be highly correlated with other PD-related metrics that have not been used in our analysis. To do so, we consider the score of activities of daily living (ADL) and two imaging biomarkers (putamen and caudate), and compute their mean values within each disease subtypes. It is apparent that all of these three external biomarkers gradually decline as the subtypes progress from normal to severe, further confirming the validity of the obtained PD subtypes.

\subsection{Reproducibility, model comparison and sensitivity analysis}
\begin{figure}
    \centering
    \subfloat[\centering Symptom grouping structure in validation datasets]
    {{\includegraphics[width=8cm]{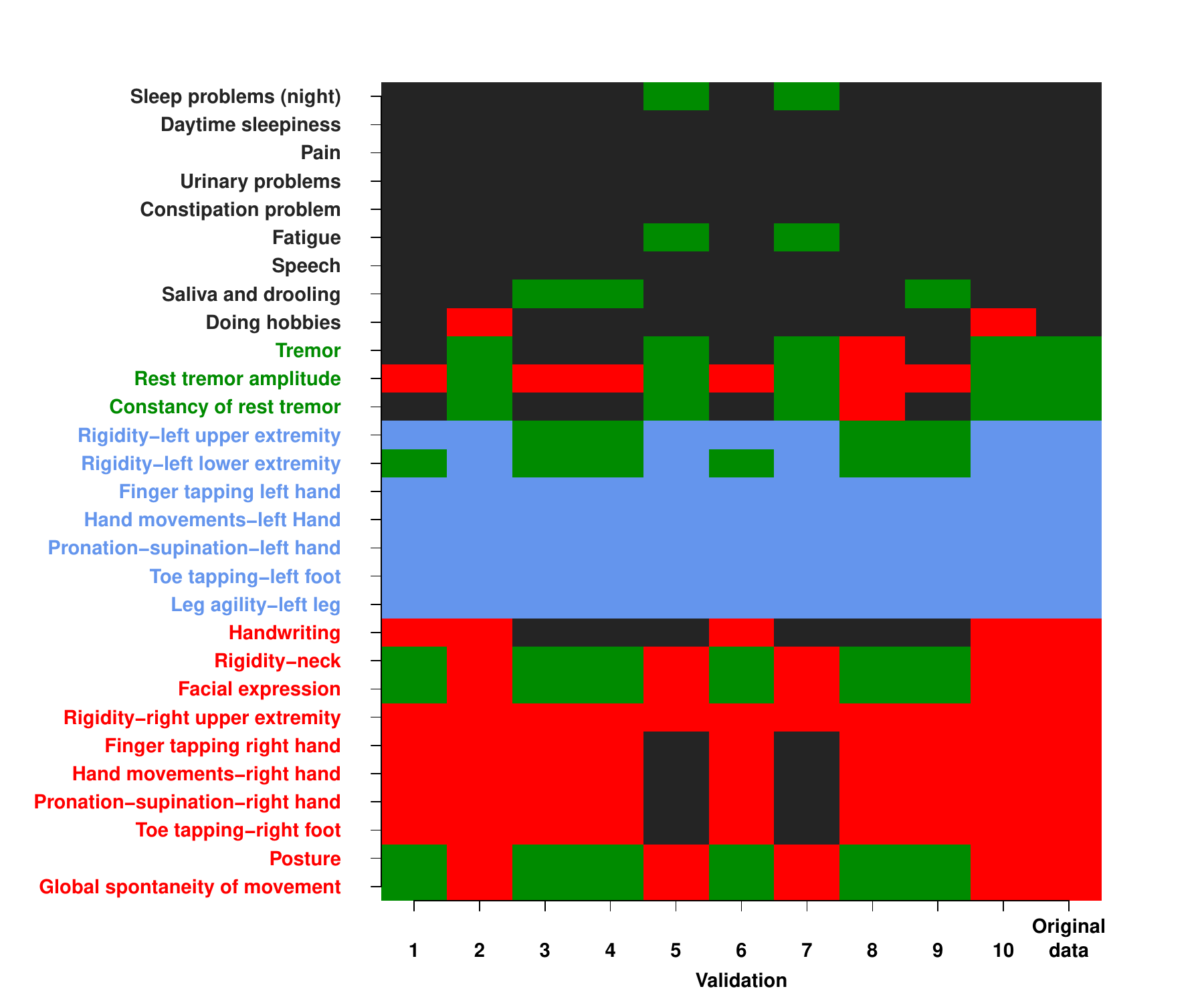} }}
    \subfloat[\centering Cut-points of the time periods in validation datasets]
    {{\includegraphics[width=8cm]{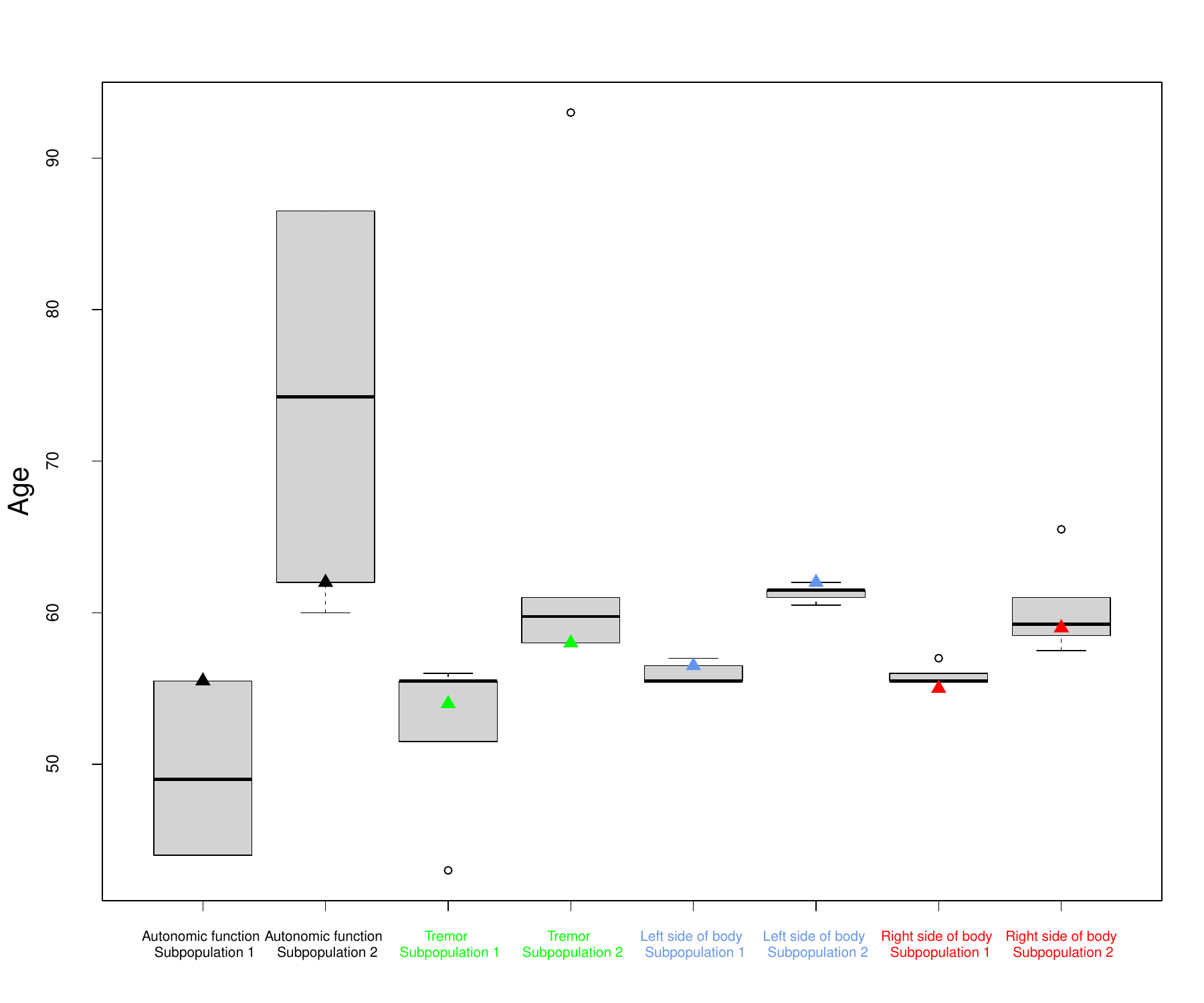} }}%
    \caption{Validation of the block structure in PPMI study. 
    Triangle symbols represent the estimates of cut-points from original dataset.}%
    \label{fig:validation}%
\end{figure}

We first evaluate the reproducibility of the block structure. Specifically, we randomly remove the longitudinal data of 10\% subjects from the dataset, treating the remaining 90\% data as a validation set. The BM$^3$ with $(G,R,K) = (4,2,3)$ is then  adopted to fit the validation set. This procedure is repeated  10 times, with each validation set containing a randomly selected 90\% subjects. Figure \ref{fig:validation} displays the symptom grouping structure and the time period cut-points  estimated from the 10 validation datasets. We observe that 18 out of the 29 MDS-UPDRS items are consistently assigned to the same symptom groups as in the original dataset in at least 80\% of the validations. Furthermore, the boxplots of the cut-points for the time periods show that, across all symptom groups and validations, the cut-points for the first subpopulation are consistently earlier than those for the second subpopulation. This observation aligns with the findings from the original dataset  (Figure \ref{pic:group_mat}(b)), confirming the existence of two subpopulations with different rates of PD progression. Additionally, the medians of the cut-points from the validation data closely match the estimates from the original data (triangle symbols in Figure \ref{fig:validation}(b)) in most symptom groups, further demonstrating the robustness and validity of the identified time periods. In conclusion, our method shows satisfactory performance in reproducing the block structure. 

\begin{figure}[h!]
            \centering
            \includegraphics[height=0.55\textheight]{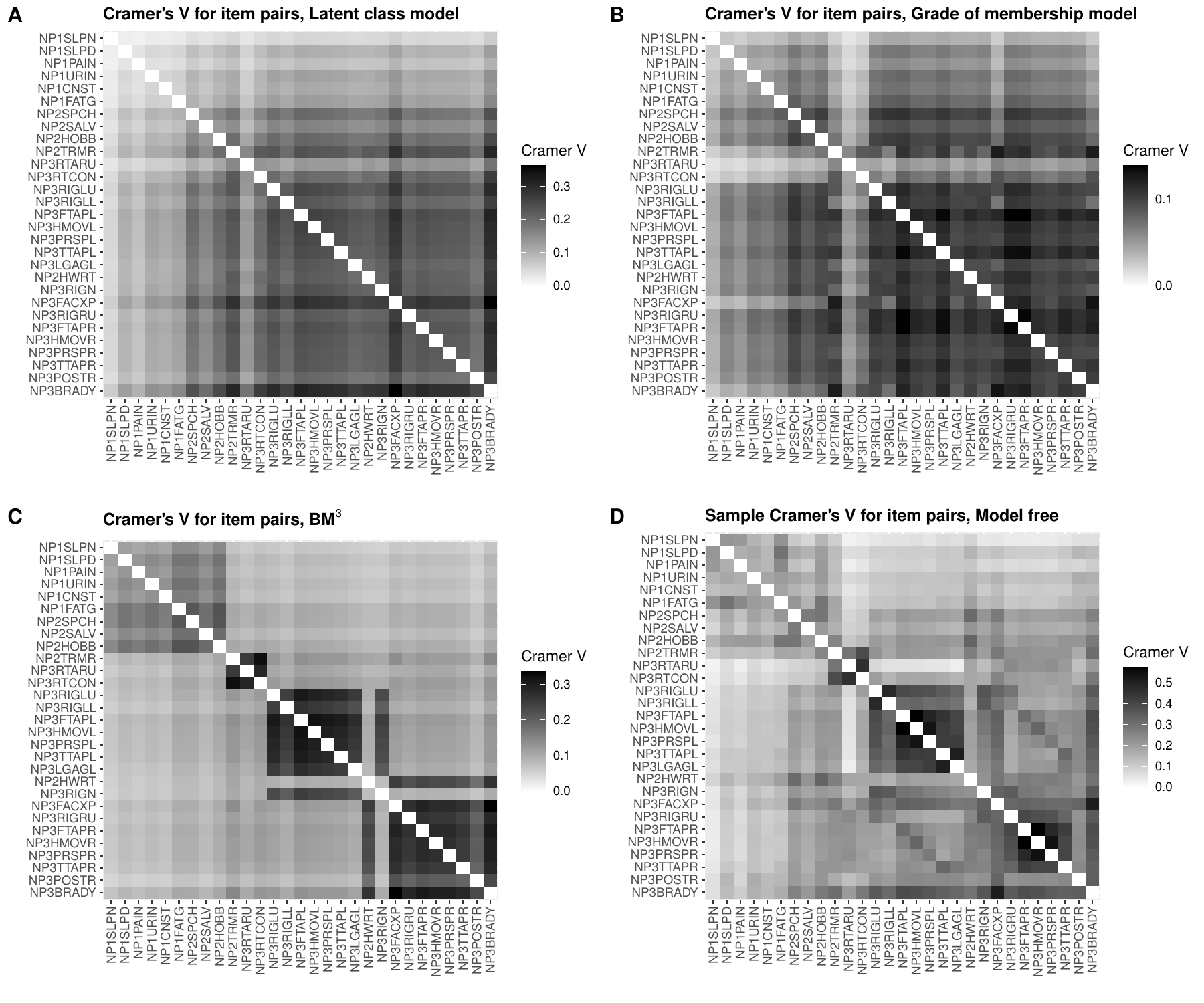}
            \caption{
            Cramer's V posterior means for item pairs obtained using latent class model, GoM and BM$^3$, and Sample Cramer’s V for item pairs calculated directly from PPMI data. } 
\label{fig:Cramer's V}
\end{figure}

For comparison, we also reanalyzed the PPMI dataset using two alternative models: the latent class model (LCM) and the conventional GoM. In the latent class model, we set the number of latent classes to 3, and for the GoM, we set the number of extreme profiles to 3.
Intuitively, the latent class model assigns each subject a single latent membership, which limits its ability to capture heterogeneous symptoms across multiple domains—for example, patients who exhibit tremor but maintain normal autonomic function. On the other hand, the GoM assigns a latent membership to each item at each time point, neglecting the potential correlations between related symptoms and adjacent time points.
To compare the performance of these models with the proposed BM$^3$, we calculated the posterior mean of Cramer's V between item pairs for each model, then compared these estimates with the sample Cramer's V values directly calculated from the PPMI data. Cramer's V is a classical measure of association between two categorical variables, which gives a value
between 0 and 1, with larger values indicating a stronger association. As shown in Figure \ref{fig:Cramer's V}, our BM$^3$ aligns more closely with the meaningful block structure of the clinical symptoms compared to the LCM and the GoM, both of which fail to adequately capture the dependence structure.

Finally, to assess the sensitivity of the Bayesian results to prior specifications, we perturb the hyperparameters $(\xi_1,\hdots,\xi_G), a_\alpha$ and $b_\alpha$ and reanalyzed the PPMI dataset. The results with disturbed priors are qualitatively similar and not presented.

\section{Simulation Studies}
We conducted Monte Carlo simulations to evaluate the performance of the proposed BM$^3$ under different block structures and sample sizes. Additionally, we assessed the ability of WAIC in selecting the number of blocks and extreme profiles.  
Based on the BM$^3$ defined in \eqref{eq:general}, we set $(p,T,G,R,K) = (10,10,2,2,4)$ and considered the following  block structure settings: \\
{\bf Setting I}: A block structure with homogeneous periods, as shown in Figure \ref{fig:block}b, where $\bs = (\bone^\top_5,2*\bone^\top_5)$ and $\bb =(\bone^\top_{5},2*\bone^\top_{5}) $, with $\bone_u$ denoting a $u$-dimensional vector of ones. \\
{\bf Setting II}: A block structure with symptom group-specific periods, as displayed in Figure \ref{fig:block}c, where $\bs = (\bone^\top_5,2*\bone^\top_5), \bb_1 = (\bone^\top_{20},2*\bone^\top_{10})$ and $\bb_2= (\bone^\top_{10},2*\bone^\top_{20})$.\\
{\bf Setting III}: A block structure with symptom group- and subpopulation-specific periods (for $C=2$ subpopulations). The true block structure is $\bs = (\bone^\top_5,2*\bone^\top_5), \bb_{1,1} =\bb_{2,1} = (\bone^\top_{3},2*\bone^\top_{7})$, and $\bb_{1,2} =\bb_{2,2} = (\bone^\top_{7},2*\bone^\top_{3})$.\\
Among the three settings of block structure, setting III is the most challenging one and mirrors the model adopted for PPMI data analysis in Section \ref{sec:realdata}.
For each setting, we considered sample sizes $n=100, 200$ and $500$, with each variable having three unordered categories ( i.e., $d_1=\hdots=d_p = 3$). The true model parameters for $\bLambda$ were specified as follows for $u=0,1$:
{\footnotesize
\begin{align}
    \bLambda_{5u+1} &= \begin{pmatrix}
0.1 & 0.30 & 0.45 & 0.70\\
0.8 & 0.10 & 0.45 & 0.05\\
0.1 & 0.60 & 0.45 & 0.25
\end{pmatrix};~\bLambda_{5u+2} = \begin{pmatrix}
0.2 & 0.45 & 0.55 & 0.80\\
0.7 & 0.05 & 0.40 & 0.10\\
0.1 & 0.05 & 0.05 & 0.10
\end{pmatrix};\nonumber\\
\bLambda_{5u+3} &= \begin{pmatrix}
0.3 & 0.45 & 0.60 & 0.90\\
0.6 & 0.05 & 0.35 & 0.05\\
0.1 & 0.50 & 0.05 & 0.55
\end{pmatrix};~\bLambda_{5u+4} = \begin{pmatrix}
0.1 & 0.25 & 0.50 & 0.90\\
0.1 & 0.65 & 0.05 & 0.05\\
0.8 & 0.10 & 0.45 & 0.05
\end{pmatrix};~
\bLambda_{5u+5} = \begin{pmatrix}
0.2 & 0.45 & 0.60 & 0.90\\
0.7 & 0.05 & 0.05 & 0.05\\
0.1 & 0.50 & 0.35 & 0.05
\end{pmatrix}.\nonumber
\end{align}}
\noindent We set the true Dirichlet parameters to $\balpha = (1,1,1,1)$ and generated 100 datasets for each setting. The proposed MCMC algorithm was used to obtain Bayesian estimates under each dataset. The hyperparameters were specified as $a_\alpha = 2, b_\alpha = 1$ and $\sigma_\alpha = 0.02$. We run three parallel MCMC chains for 10000 iterations with different initial values, and after checking the traceplots, we discarded the first 5,000 iterations as burn-in. For continuous parameters $\bLambda$ and $\balpha$, we computed the posterior mean, while the posterior modes of block indicators $\bs, \bb$ and latent allocation variables $z_{i,j,t}$ were calculated. The Adjusted Rand Index (ARI) \citep{rand1971objective}, a measure of similarity between two clusterings, was used to compare the estimated $\bs$ and $\bb$ with the true values. The ARI ranges from 0 to 1, with 1 indicating perfect agreement. We also computed the Root Mean Squared Errors (RMSEs) for $\bLambda$ and $\balpha$ to assess estimation accuracy. Table \ref{tab: simu all} summarizes the estimation results across different block structures and sample sizes. The ARIs for symptom grouping $\bs$ and time grouping $\bb$ are quite high across all settings. As sample size $n$ increased, the RMSEs for $\bLambda$ and $\balpha$ decreased, indicating improved estimation accuracy. Overall, the Bayesian estimates showed satisfactory accuracy in the simulations, confirming the identifiability and posterior consistency of the model parameters in BM$^3$.

\begin{table}[ht]
\caption{Parameter estimates of BM$^3$ for $(p,T,G,R,K) = (10,10,2,2,4)$ with homogeneous and heterogeneous periods.}
\label{tab: simu all}
\centering
\scriptsize
\resizebox{\columnwidth}{!}{%
\begin{tabular}{cccccccccccc}
\hline
   \multirow{2}{*}{n}     & \multicolumn{2}{c}{ARI of $\bs$} &  & \multicolumn{2}{c}{ARI of $\bb$} &  & \multicolumn{2}{c}{RMSE of $\bLambda$} &  & \multicolumn{2}{c}{RMSE of $\balpha$} 
           \\ \cline{2-3} \cline{5-6} \cline{8-9} \cline{11-12} 
        & \multicolumn{1}{c}{Median}  & \multicolumn{1}{c}{(IQR)}& &   \multicolumn{1}{c}{Median}  & \multicolumn{1}{c}{(IQR)}& &   \multicolumn{1}{c}{Median}  & \multicolumn{1}{c}{(IQR)}& &   \multicolumn{1}{c}{Median}  & \multicolumn{1}{c}{(IQR)}
\\ \hline
\multicolumn{12}{c}{Homogeneous periods $\bb$ }\\ \cline{1-12}
100  & 1.00              & (0.00)             &  & 1.00              & (0.00)               &  & 0.077              & (0.021)    &  & 0.212              & (0.474)\\ 

    200  & 1.00              & (0.00)             &  & 1.00              & (0.00)               &  & 0.036              & (0.010)    &  & 0.073              & (0.161)\\    
500  & 1.00              & (0.00)             &  & 1.00              & (0.00)               &  & 0.015              & (0.004)    &  & 0.034              & (0.063)\\ \hline
\multicolumn{12}{c}{Symptom group-specific periods $\bb_g$ }\\ \cline{1-12}
 100  & 1.00              & (0.00)             &  & 1.00              & (0.00)               &  & 0.079             & (0.031)    &  & 0.193              & (0.374)\\

200  & 1.00              & (0.00)             &  & 1.00              & (0.00)               &  & 0.037              & (0.008)    &  & 0.100              & (0.186)\\
500  & 1.00              & (0.00)             &  & 1.00              & (0.00)               &  & 0.016             & (0.006)    &  & 0.035             & (0.112)\\ \hline
\multicolumn{12}{c}{Symptom group- and subpopulation-specific periods $\bb_{g,c}$}\\ \cline{1-12}
 100  & 1.00              & (0.00)             &  & 1.00              & (0.00)               &  & 0.080              & (0.025)    &  & 0.176             & (0.491)\\ 

200  & 1.00              & (0.00)             &  & 1.00              & (0.00)               &  & 0.036              & (0.009)    &  & 0.089              & (0.203)\\ 
   
500  & 1.00              & (0.00)             &  & 1.00              & (0.00)               &  & 0.015              & (0.006)    &  & 0.039              & (0.071)\\
\hline

\end{tabular}%
}
\end{table}

To evaluate the performance of WAIC in determining the number of extreme profiles and block structures in BM$^3$, we carry out a simulation study by generating 100 datasets with $(n,p,T,G,R,K)=(100,15,10,3,2,4)$. The true population values for parameters are set as the same as in the parameter estimation simulations. To mimic the scenario in PPMI data analysis, we fixed $R=2$ here and considered competing models with $G=2,3,4$ and $K=3,4,5$, leading to 9 competing models. The corresponding WAIC values were calculated for each model. When fixing the candidate $G$ to the truth $G=3$ and varying $K=3,4,5$,  the percentages of the datasets that
each of $K = 2, 3, 4, 5$ is selected are 3\%, 85\% (true $K$), 12\%, respectively. When fixing the candidate $K$ to the truth $K=4$ and varying $G=2,3,4$,  the percentages of the datasets that
each of $G = 2, 3, 4$ is selected are 0\%, 57\%(true $G$), 43\%, respectively. Additionally, the mean WAIC values of $G=2, 3, 4$ are 37,161, 23,379 and 23,394, respectively.  In conclusion, WAIC does not tend to underestimate $G$ and $K$, and that it has a reasonably good accuracy of selecting the
truth.   Further, we recommend to choose a more parsimonious model if the WAIC difference is small (e.g., less than 10).

\section{Discussion}
Motivated by the observation that patients affected by PD exhibit substantial between-individual, between-symptom, and between-time heterogeneity during the disease progression,
we have applied a novel blockwise mixed membership model that partitions symptoms and time points into several blocks, assigning a latent membership to each block. Our model allows each subject to belong to multiple latent clusters and capture the different disease statuses across various symptom domains and time periods. In our analysis of the PPMI study, which included 29 PD-related clinical symptoms collected from 1,531 subjects over a maximum of 18 follow-up visits, we aimed to discover the clinical heterogeneity of PD during disease course and identify PD subtypes. Our findings revealed four clinically meaningful symptom domains, two distinct time periods, and three extreme disease profiles representing different levels of dysfunction in autonomic function, tremor, and motor function. We validated disease subtypes using external variables and found that they aligned with patterns in medication initiation, neuroimaging biomarkers, and activities of daily living. Moreover, our model outperformed conventional GoMs and latent class models in capturing correlations among PD-related symptoms.

Several extensions of this work are worth considering. First, the BM$^3$ could be extended to handle continuous variables. In addition to categorical assessments, the PPMI study collects various continuous data, such as cerebrospinal fluid measurements, neuroimaging results, and microRNA biomarkers. However, the identifiability of such a more complex model remains uncertain and warrants further investigation. Second, while we used the WAIC to select the number of extreme profiles, symptom groups, and time periods, this pairwise model selection process becomes increasingly tedious and computationally demanding as the model space expands. A promising future direction would be to develop a simultaneous parameter estimation and model selection procedure, streamlining the process.
Third, incorporating covariates (e.g., gender, genetic mutations) into the modeling of the conditional probability $\blambda$ and/or the Dirichlet parameter $\balpha$ could be highly informative. This extension would allow us to identify extreme profile-specific covariate effects on MDS-UPDRS items, potentially providing deeper insights into PD progression.

\section*{Supplementary Material}
The Supplementary Material contains the proofs of the theoretical results and additional computation details.

\spacingset{1}
\bibliographystyle{apalike}
\bibliography{bib_template}

\end{document}